\documentclass[10pt]{sigplanconf-eurosys}

\usepackage{multirow}
\usepackage{amssymb,amsmath}
\usepackage{graphicx} 
\usepackage{subfig}
\usepackage{algorithm}
\usepackage{algorithmic}
\usepackage{color, soul}
\soulregister\cite7 
\soulregister\ref7
\soulregister\pageref7
\usepackage{array}
\usepackage{cases}
\usepackage{url}
\usepackage{colortbl}

\usepackage{multicol}
\usepackage{subfloat}

\definecolor{Gray}{gray}{0.9}

\newcounter{MANumberOfComments}
\stepcounter{MANumberOfComments}
\newcounter{LINumberOfComments}
\stepcounter{LINumberOfComments}
\newcounter{JBNumberOfComments}
\stepcounter{JBNumberOfComments}
\newcounter{ReviewComments}
\stepcounter{ReviewComments}
\newcounter{TysonComments}
\stepcounter{TysonComments}

\iftrue
\newcommand{\rone}[1]{}
\newcommand{\rtwo}[1]{}
\newcommand{\rthree}[1]{}
\newcommand{\liang}[1]{}
\newcommand{\manote}[1]{}
\newcommand{\jbnote}[1]{}
\newcommand{\tyson}[1]{}
\fi

\usepackage{amssymb}
\usepackage{pifont}
\newcommand{\cmark}{\textcolor{green}{\ding{51}}}%
\newcommand{\xmark}{\textcolor{red}{\ding{55}}}%
\usepackage{booktabs}

\usepackage{etoolbox}
\makeatletter
\patchcmd{\maketitle}{\@copyrightspace}{}{}{}
\makeatother

\begin{document}


\setlength{\pdfpageheight}{\paperheight}
\setlength{\pdfpagewidth}{\paperwidth}

\title{Diffusing Your Mobile Apps: Extending In-Network Function Virtualization to Mobile Function Offloading}



\authorinfo{M\'{a}rio Almeida}{Universitat Polit\`{e}cnica de Catalunya}{4knahs@gmail.com}
\authorinfo{Liang Wang}{University of Cambridge}{liang.wang@cl.cam.ac.uk }
\authorinfo{Jeremy Blackburn}{Telefonica Research}{jeremy.blackburn@gmail.com }
\authorinfo{Konstantina Papagiannaki}{Telefonica Research}{dina@tid.es }
\authorinfo{Jon Crowcroft}{University of Cambridge}{Jon.Crowcroft@cl.cam.ac.uk }


\maketitle

\begin{abstract} 

Motivated by the huge disparity between the limited battery capacity of user devices and the ever-growing energy demands of modern mobile apps, we propose INFv.
It is the first offloading system able to cache, migrate and dynamically execute on demand functionality from mobile devices in ISP networks.
It aims to bridge this gap by extending the promising NFV paradigm to mobile applications in order to exploit in-network resources.


In this paper, we present the overall design, state-of-the-art technologies adopted, and various engineering details in the INFv system. We also carefully study the deployment configurations by investigating over 20K Google Play apps, as well as thorough evaluations with realistic settings. 
In addition to a significant improvement in battery life (up to 6.9x energy reduction) and execution time (up to 4x faster), INFv has two distinct advantages over previous systems: 1) a non-intrusive offloading mechanism transparent to existing apps; 2) an inherent framework support to effectively balance computation load and exploit the proximity of in-network resources. Both advantages together enable a scalable and incremental deployment of computation offloading framework in practical ISPs' networks. 


\end{abstract}

\section{Introduction}
\label{sec:intro}


Pervasive mobile clients have given birth to complex mobile apps, many of which require a significant amount of computational power on users' devices.
Unfortunately, given current battery technology, these demanding apps impose a huge burden on energy constrained devices. 
While power hogging apps are responsible for 41\% degradation of battery life on average~\cite{oliner_carat:_2013}, even popular ones such as social networks and instant messaging apps (e.g., Facebook and Skype) can drain a device's battery up to \emph{nine times faster} due only to maintaining an on-line presence~\cite{Aucinas:2013:SOW:2535372.2535408}.

\begin{table*}[t!]
\centering
\begin{tabular}{@{}cccccccc@{}}
\toprule
\textbf{MCO}          & \textbf{Partitions} & \textbf{Dynamic} & \textbf{No Repackage} & \textbf{Stock OS}      & \textbf{Cloud} & \textbf{Network} & \textbf{Deployment} \\ \midrule
MAUI~\cite{cuervo_maui:_2010}         & Manual / Method         & \xmark  & \xmark       & \cmark & \cmark     & \xmark        & \xmark     \\
ThinkAir~\cite{kosta_thinkair:_2012}     & Manual / Method          & \xmark  & \xmark       & \cmark        & \cmark     & \xmark        & \cmark (EC2,cost)     \\
CloneCloud~\cite{chun_clonecloud:_2011}   & Auto / Thread          & \cmark  & \xmark       & \xmark        & \cmark     & \xmark        & \xmark     \\
Comet~\cite{gordon_comet:_2012}        & Auto / Thread           & \textbf{--}  & \cmark       & \xmark        & \cmark     & \xmark        & \xmark     \\
Zhang et al.~\cite{zhang_refactoring_2012} & Auto / Class              & \xmark  & \xmark       & \cmark        & \cmark     & \xmark        & \xmark     \\
INFv         & Auto / Class              & \cmark  & \cmark       & \cmark        & \cmark     & \cmark        & \cmark (cache,load)    \\ \bottomrule
\end{tabular}
\caption{Mobile Cloud Offloading~(MCO) systems and properties. The comparison reveals that INFv supersedes the previous designs in many aspects. (red cross means the feature is not supported whereas green tick means the opposite.)}
\label{tab:infvrules}
\vspace{-5mm}
\end{table*}

Recent work has proposed various solutions to offload and execute functionality of mobile apps remotely in a cloud, referred to as a mobile-to-cloud paradigm~\cite{cuervo_maui:_2010,kemp_cuckoo:_2010,kosta_thinkair:_2012,chen_coca:_2012, zhang_refactoring_2012,gordon_comet:_2012,chun_clonecloud:_2011}. 
Their evaluations have shown that the energy consumption of CPU intensive apps, e.g., multimedia processing apps and video games, can be reduced by an order of magnitude~\cite{chun_clonecloud:_2011, cuervo_maui:_2010}. 
Beside the extended battery life, there are other benefits, such as faster execution time, responsiveness, and enhanced security by dynamic patching~\cite{mulliner_patchdroid:_2013}.

However, prior work suffers from two limitations. 
First, they overlooked the potential of exploiting ISPs' in-network resources for functionality offloading, simply using the network as a transmission fabric.
Quite different from a decade ago, network middle boxes are no longer simple devices which only forward packets, often featuring multi-core general purpose processors~\cite{xeon} far more capable than those of mobile devices. 
In fact, many ISPs' own network services have been shifting from specialized servers to generic hardware with the adoption of the NFV (Network Function Virtualization) paradigm\footnote{Telefonica aimed to shift 30\% of their infrastructure to NFV technologies by the end of 2016~\cite{telefonica:nfv,telefonica:cartablanco,telefonica:unica}. 
Other providers such as AT\&T~\cite{att:nfv}, Vodafone~\cite{ericsson:nfv}, NTT Docomo~\cite{nttdocomo:nfv} and China Mobile~\cite{chinamobile:nfv} are following similar steps.}. 
This paradigm can be naturally extended from basic network functions (e.g., packet filtering) to the more general functionality of mobile apps,
exploiting ``last-hop'' proximity to effectively reduce latency, network load, and improve availability compared to a centralized mobile-to-cloud deployment. 
When deployed close to cellular towers (Radio Access Network), offloading latency could be reduced by up to 86.7\%, reducing the energy consumption and execution time by up to 21.6\% and 24.5\%, respectively, when compared to a popular cloud alternative (Section~\ref{sec:runtimeeval}). 
Furthermore, such a system could potentially be extended to adapt an app's lifecycle to network conditions, further reducing devices' energy consumption (e.g., delay network dependent background execution~\cite{almeida_empirical_2016} in the case of congestion) and the volume of signaling offloaded to the Core Network (CN)~\cite{patel2014mobile}. 
Unfortunately, previous systems either failed to address the challenges of deploying and scaling mobile code offloading systems at all, or overlooked the opportunities to effectively exploit in-network resources.
Second, these solutions often utilize intrusive offloading techniques which either require custom OS distributions~\cite{chun_clonecloud:_2011,gordon_comet:_2012}, app repackaging~\cite{zhang_refactoring_2012}, or even alternative app stores, not only increasing security risks and deployment costs, but also greatly increasing the barrier to the market adoption.

\rthree{- the merits of operator offload are not obvious (since wired latencies are very small when compared to wireless ones) - the authors should quantify them better.
- it is not obvious to what extent the design leverages operator networks (or their proximity to the mobile users) - all the network part is done in simulation.  you fail to convince the reader there is a single benefit that can be had in operator networks that cannot be had in a cloud offloading scenarios (all major cloud operators have presence in all continents and wired latencies should be well below 50ms - so why should we care about operator offload?).}

\rthree{you need a crisp reason for why operator offload makes sense (aside from the operator making some cash), and you need to implement a full system that actually relies on operator deployments to get its job done.}

\manote{I think I briefly addressed these complaints in MA\#1 by adding the latency, energy and execution time numbers retrieved from our last experiment.}

This paper presents INFv, the first mobile computation offloading system able to cache, migrate, and dynamically execute mobile app functionality on demand in an ISP network. 
It uses advanced interception and automatic app partitioning based on functionality clustering, combined with in-network load balancing algorithms to perform transparent, non-intrusive, in-network functionality offloading. 
INFv aims to bridge the gap between the limited battery capacity of user devices and the ever-growing energy demands of modern mobile apps\cite{oliner_carat:_2013,Aucinas:2013:SOW:2535372.2535408} by extending the promising NFV paradigm to mobile apps. 
More specifically, we make the following contributions:

1)~We present INFv's data-driven architecture and design, along with key mechanisms and various technical details required to achieve non-intrusive offloading and adaptive in-network resource management.

2)~We show that INFv is able to greatly improve apps energy consumption (reduced by up to 6.9x) and speed up app execution (up to 4x faster). It performs similar to, or better than, local execution 93.2\% of the time over 4G while adapting to dynamic network conditions and up to 24.5\% faster than a cloud alternative.

3)~We compare different strategies to effectively balance functionality load in the network while reducing both end-user-experienced latency and request drop rates.

4)~Through a real mobile app market study we show that app's storage cost can be reduced by up to 93.5\%, and that top apps have a median of 17 distinct functionality clusters, with up to 57\% of offloadable code.



\section{Background \& Related Work}
\label{sec:related}

\rthree{please have a look at Mobile Edge Computing (MEC) which may be relevant to what you are trying to achieve.}
\tyson{Sounds like you might want to include more stuff on RPC? Particularly work on transparent RPC used for offloading?}



Mobile Code Offloading~(MCO) is a reasonably well explored area~(see Table~\ref{tab:infvrules}), however earlier work has a few limitations that we directly address with INFv.
In this section we provide an overview of previous work, focusing on the lessons taken away that we used to design INFv.
MCO systems can be differentiated based on their granularity and partitioning decisions (what to offload), offloading techniques (how to offload), and runtime decisions (when to offload).

\textbf{What to offload} can be defined in a \emph{manual} (app developer assisted) or in an \emph{automated} manner.
The first can be accomplished via programming frameworks and/or code annotations.
For programming frameworks~\cite{kemp_cuckoo:_2010,chen_coca:_2012,kosta_thinkair:_2012}, both local and remote execution alternatives have to be implemented according to the framework's design constraints (e.g., concurrency models).
In Maui~\cite{cuervo_maui:_2010}, annotations allow a partially automated offloading solution where developers select methods to offload.
The benefit of these explicit systems is the level of customization; developers have a large degree of control over how their apps are offloaded.

An alternative approach taken by other work is to make automated offloading decisions~\cite{chun_clonecloud:_2011,gordon_comet:_2012,zhang_refactoring_2012} by performing static and dynamic analysis of apps.
While automated approaches give up a degree of flexibility, they benefit from being able to leverage the existing app ecosystem and general ease of use.
That said, these systems do not really focus on the deployment characteristics of offloading.
Thinkair~\cite{kosta_thinkair:_2012} and Cloudlets~\cite{satyanarayanan_case_2009}, however do to some extent.
Thinkair allows on-demand execution in a cloud environment.
It provides 6 different VM types, with varying CPU and memory configurations.
Mobile devices upload \emph{specially crafted apps} to the cloud, and their local counterpart negotiates the on-demand execution on one of these VMs.
A more robust approach (one that we have taken) is to support \emph{existing apps} while handling resource negotiation and functionality caching in a fully automated and transparent manner.
In particular, we want to ensure that INFv supports heterogeneous network topologies and load balances cached functionality in an intelligent manner.
Cloudlets~\cite{satyanarayanan_case_2009} in particular serves as a motivation for INFv as \emph{it highlights the impact of high latency in MCO} to justify the need for deploying physically proximate decentralized clusters to execute functionality.
INFv directly addresses the problems and challenges raised in this work by proposing an in-network solution (``Network'' in Table~\ref{tab:infvrules}) for MCO.

The majority of offloading literature proposes a method offloading granularity~\cite{cuervo_maui:_2010,kemp_cuckoo:_2010,kosta_thinkair:_2012,chen_coca:_2012}.
CloneCloud~\cite{chun_clonecloud:_2011} and Comet~\cite{gordon_comet:_2012} propose full or partial thread granularity.
Automated method granularity architectures incur the cost of synchronizing the serialized method caller objects, parameters, changed state and return object.
Thread granularity architectures often need to synchronize thread state, virtual state, program counters, registers, stack or locks. 
INFv is the \emph{first to address the challenges behind caching app functionality, complementing its granularity design with a real market study to reduce its app storage requirements.}



\textbf{How to offload} depends on the aforementioned characteristics.
Manual approaches tend to use custom compilers (e.g., AspectJ) or builders.
Automated solutions often operate on compiled apps and rely on byte-code rewriting~\cite{zhang_refactoring_2012} or VM modifications~\cite{chun_clonecloud:_2011,gordon_comet:_2012}. 
Altering an app generally requires repackaging and resigning it, which also implies the need for a new distribution mechanism incompatible with current app markets.
Each of the above architectures, except for Comet~\cite{gordon_comet:_2012} (a distributed shared memory solution), need either app repackaging, re-writes, or both, impacting their likelihood of adoption.
INFv offloading differs mainly in that it \emph{does not require a custom distribution, manual intervention or app-repackaging even in the presence of app updates.}
It allows for dynamically loaded partitions (``Dynamic'' in Table~\ref{tab:infvrules}) and is fully reversible.


\textbf{When to offload} was mostly done using thresholds~\cite{gordon_comet:_2012} or Integer Linear Programming~\cite{cuervo_maui:_2010,chun_clonecloud:_2011} based on the app profiling metrics.
Similar to CloneCloud, INFv relies on UI instrumentation to profile the different execution paths of apps.
INFv performs app profiling on remote servers to reduce the overhead on mobile devices and like Thinkair~\cite{kosta_thinkair:_2012}, its energy model is based on PowerTutor~\cite{zhang_accurate_2010}.
INFv improves previous systems by \emph{offloading together functionality with high communication based on their runtime invocation frequency}.  

\textbf{Clone detection} literature~\cite{mojica_large-scale_2014,linares-vasquez_revisiting_2014,ruiz_understanding_2012,desnos_android:_2012} focused on detecting app cloning using class/method names and tend to ignore minor custom changes to the functionality.
In code caching, we are more interested in the unmodified use of third-party libraries than similar code.
Unlike recent studies~\cite{linares-vasquez_revisiting_2014,mojica_large-scale_2014,wang_wukong:_2015}, some initial works~\cite{ruiz_understanding_2012, desnos_android:_2012} did not consider obfuscation, which can impact the statistical significance of their results.
In~\cite{mojica_large-scale_2014} and~\cite{linares-vasquez_revisiting_2014} the authors study obfuscation based on class names. 
Unfortunately, 
they \emph{do not consider package obfuscation which affects the offloading routing mechanisms}.
Wukong et al.~\cite{wang_wukong:_2015} focused on clone detection based on Android API calls, and while it highlights the challenges in overcoming obfuscation, we have found strong evidence that, \emph{a simple package name filtering might suffice to detect obfuscation at a package level.}

\textbf{Mobile Edge Computing}~\cite{mec} (MEC) is an industry initiative\footnote{supported by Huawei, IBM, Intel, Nokia, NTT DOCOMO and Vodafone.} whose goal is to provide computing capabilities at the edge of the cellular network.
Its focus is explicitly on the infrastructure and deployment and not on the potential applications.
That said, INFv can be considered an obvious use case of MEC and the first full-fledged MCO architecture to exploit its potential.

\textbf{Runtime patching} has been used to dynamically provide updates to apps.
OPUS~\cite{altekar_opus:_2005} focused on providing dynamic software patching to C programs, while POLUS~\cite{chen_polus:_2007} was more focused on updating long-lived server side apps.
More recently, such techniques were brought to Android with PatchDroid~\cite{mulliner_patchdroid:_2013}.
It focused on security vulnerabilities and proposed a system to distribute and apply third-party in-memory security patches. 
Inspired by these systems, INFv modifies a single Android OS binary to extends app's functionality \emph{at runtime}, providing mechanisms similar to those of aspect oriented programming.

\section{Design Goals \& Architecture}
\label{sec:arch}

\begin{figure}
  \centering
    \includegraphics[width=0.45\textwidth]{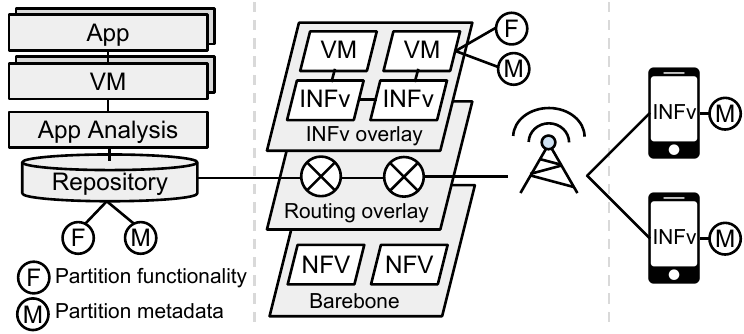}
    \caption{Design of INFv overall system, the three subsystems are divided with gray dashed lines.}
    \label{fig:architecture}
    \vspace{-5mm}
\end{figure}


Based on the limitations of previous work, and keeping in mind the recent availability of in-network resources, we identified four major design requirements for INFv as well as their corresponding challenges.

\textbf{Instrumenting apps:} App instrumentation is needed for profiling and providing offloading capabilities. How can apps be modified without performing app repackaging, using specialized compilers, custom OS or app stores? 

\textbf{Understanding apps:} With instrumentation in place, how to detect which functionality is consuming the most energy or executing for longer? Can we analyze apps without incurring a performance penalty on mobile devices?

\textbf{Offloading functionality:} How to enable apps to make use of remote resources in a transparent and efficient manner with minimal adoption cost? Can offloading adapt to a dynamic network environment and still ensure its energy and performance benefits? 

\textbf{Caching functionality:} How to cache functionality in a network, adapting to demand and reducing latency, even for arbitrary network topologies and heterogeneous nodes?

\textbf{Deployment assumptions:} Based on recent/expected NFV/MEC adoption by ISPs, we believe these standards will soon be widely available in the ISPs networks.
More specifically, and in compliance with the MEC proposal, placed in the Radio Access Network (RAN), where traffic offloading functions can be implemented to filter packets based on their end-point~\cite{mec,lipa} (IP protocol).

The overall INFv architecture that addresses these challenges is depicted in Fig.~\ref{fig:architecture}.
It is divided in three different logical subsystems (separated by dashed lines), each addressing the aforementioned challenges.
The leftmost subsystem (Section~\ref{sec:profiling}) profiles and analyzes apps. 
The rightmost subsystem (Section~\ref{sec:offloading}) provides the on device offloading capabilities.
It runs on the user device and executes code on a remote VM in the network.
Finally, the third subsystem (Section~\ref{sec:network}) runs on the network nodes (NFV) and is responsible for caching functionality and balancing the computation load.
In the next sections we detail each of them and describe their technical challenges.

\subsection{Dynamic instrumentation}
\label{sec:instrumentation}


INFv's mechanism to extend app functionality has to have minimal impact and high coverage of devices, i.e., it cannot rely on app repackaging, specialized compilers, custom OS or app stores.
To avoid modifying app binaries and change apps' signatures, INFv targets the minimal set of changes required to provide dynamic instrumentation -- a reversible binary patch to Android's \textit{app\_process}.
This process is launched at boot (by the \textit{init.rc} script) and launches Zygote -- Android's daemon responsible for launching apps, creating the Dalvik VM and pre-loading Java classes.
When a new app is to be launched, this process is forked and the app executes on its own VM with its copy of the systems libraries.
By extending this process INFv can add its own classes to the classpath and redirect method invocations to a generic redirection method that allows specifying methods to be intercepted (i.e., hooks).

\begin{figure}
  \centering
    \includegraphics[width=0.35\textwidth]{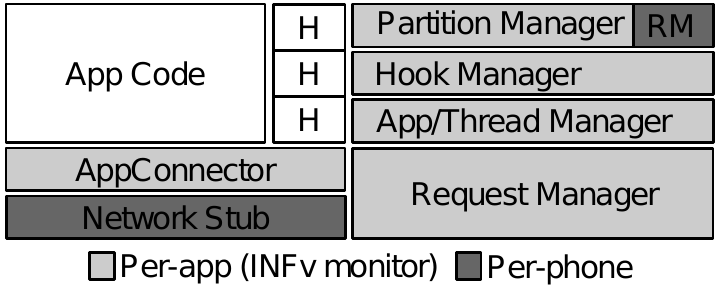}
    \caption{INFv mobile architecture. All components except the Network Stub and the Resource Manager (RM) are contained within the app process.}
    \label{fig:mobile}
    \vspace{-6mm}
\end{figure}

There are a few dynamic instrumentation frameworks available for Android, that follow similar techniques, such as Xposed~\cite{xposed}, Cydia~\cite{cydia} and adbi~\cite{adbi}.
INFv is based on the first as it is by far the most popular.
As shown in Fig.~\ref{fig:mobile}, INFv extends it to enable app profiling~(Resource Manager -- RM), manipulate app lifecycle (App/Thread Manager), perform remote method invocation (Hook Manager and the generic offloading hook -- H) and instantiates a custom classloader (\textit{DexClassLoader}) to dynamically load only its signed application strategies and functionality in the backend (Partition Manager). 
These are detailed in the next sections and their limitations and security concerns discussed in Section~\ref{sec:discussion}.

\subsection{Profiling and Partitions}
\label{sec:profiling}

In MCO, apps are typically \emph{partitioned} into a set of functionality to execute locally and another to execute remotely.
Such systems need to detect good candidates to offload, e.g., based on energy consumption and execution time.
Next, we describe how INFv partitions are designed and how the environment-dependent properties (e.g., network conditions) are considered for offloading decisions.



\subsubsection{App Analysis \& Profiling}
\label{sec:analysis}

Designing functionality partitions requires app knowledge. 
In INFv most of app analysis is done offline (left-most system in Fig~\ref{fig:architecture}) and is composed of two steps.
The first consists of performing static analysis of the app, i.e., the app analysis platform retrieves an app from the Google Play (GP) store, decompiles it (using \textit{apktool}\cite{apktool}) and parses both the manifest and \textit{smali} files (i.e., disassembled Dalvik bytecode).
From the first we retrieve the app properties, such as package name, version and app starting points (e.g., activities, broadcast receivers, etc).
From the second we build a call graph where the vertexes represent classes and the edges method calls.
This call graph is used to detect accesses to methods/classes that should execute locally, such as access to local hardware~\footnote{Android's bluetooth, hardware, MTP, NFC and telephony packages.} and UI~\footnote{Android's Activity, View, Widget, Gesture, Text, Transition, Animation and Graphics packages.}.
Our static analysis tool~\footnote{We made our static analysis tools open-source~\cite{broker,droidsmali}.} includes a DSL to tag packages which we used to manually pre-tag most of the Android packages depending on their requirements.

The second step consists of attributing weights to the call graph edges based on their runtime invocation counts as well as metadata regarding vertices' energy consumptions and execution times.
Apps are executed in instrumented VMs and exercised using UI automation (reproducible pseudo-random input~\cite{monkey}).
These VMs use the aforementioned instrumentation mechanisms to load the call graph on app start and intercept method invocations.
We had to perform some optimizations to reduce the tracing overhead, such as increasing the VM heap size and restricting the tracing to app methods.
The average of the intercepted methods parameters are registered in the call graph metadata -- state size, thread, execution time, invocation counts -- along with the predicted energy consumption based on the PowerTutor model~\cite{zhang_accurate_2010}.
This model uses the device states (CPU, screen, GPS, connectivity) to predict energy consumption.
It can derive a per device power model (with a low error of up to 2.5\%) by performing regression over the energy discharge patterns observed while looping through different device power states.
The INFv dynamic analysis platform can accommodate physical devices to automate this process, or a one time power discharge execution can be run in users devices to retrieve the specific device model.
We report that the minimal information needed by our offloading mechanisms requires on average 3MB per app when uncompressed (based on the top 30 market apps), which is reasonable even considering Android's memory constraints (i.e., small heap size -- 48MB in Galaxy S2).

While INFv's approach saves the profiling energy on mobile devices as well as the overhead of method tracing, we do acknowledge some limitations.
Some of the power model metrics (e.g., screen, GPS) are not accurate or observable in a VM.
The VM screen is always on and some networks/sensors are emulated.
While the second is unimportant as these should execute locally anyway, the first may overestimate energy consumption which later prevents some functionality from being offloaded.
We discuss some of these limitations in Section~\ref{sec:discussion}.

\subsubsection{App Partitions}
\label{sec:partitions}

Partitions define which subsets of functionality should be offloaded.
In MCO, partitions are generally static and their outcome is a rebuilt app (potentially two disjoint apps) with some functionality re-purposed for offloading.
In INFv's client, partitions are just information regarding the offloading subset, i.e., class names and execution properties (e.g., execution time, energy consumption, network conditions).
These sets are pushed to the device, loaded on app launch and drive app's execution.

Partitions can be devised on method and thread granularity.
The latter incurs an extra cost of synchronization (e.g., thread state, virtual state, program counters, registers, stack), and is often restricted to method boundaries~\cite{chun_clonecloud:_2011}.
In order to better integrate with the NFV abstraction, our solution \emph{is} able to provide the granularity of method offloading. 
However, since most mobile architectures apps are developed in class-based object-oriented languages, in practice we use a class offloading granularity as methods often invoke methods of the same class and share class state (e.g., class fields).
Furthermore, there is a large overlap of classes and packages across apps, which minimizes the impact of storing app functionality in the network (Section~\ref{sec:cache}). 

One of the main concerns in MCO is the balance between offloading computation and its communication overhead due to remote method invocations.
To reduce this overhead, we use the app call graph (Section~\ref{sec:analysis}), i.e., $G=(V,E)$ where $V$ is the set of app classes and $E$ the invocations between classes, to detect communities/clusters (sub-graphs of $G$) of $V$ based on their invocation counts. 
INFv uses a community detection algorithm -- Girvan and Newman (GN) \cite{girvan_community_2002} to detect edges that are likely between communities.
To do so it recursively detects the edge with highest number of shortest paths between pairs of nodes passing through it (i.e., the edge with highest betweenness) and removes it.
By removing edges with high betweenness, the communities get separated and we end up with distinct functionality clusters.
Community detection was firstly proposed by Zhang et al.~\cite{zhang_refactoring_2012} which used static analysis invocations as edge weights.
The problem is that these do not reflect the actual runtime invocation counts between classes, and so the study relies on a weight heuristic based on class semantic similarity, i.e., class names and their textual contents.
Unfortunately, our app study in Section~\ref{sec:cache} indicates that 82\% of the apps perform class name obfuscation and in some cases even the strings within classes can be obfuscated~\cite{dexguard}.
As INFv executes apps, it registers the method's runtime invocation counts, bypassing such limitations.

The GN algorithm receives the number of clusters as a parameter which we iterate from two to the optimal number of clusters as calculated via louvain's~\cite{blondel2008fast} modularity (density of links within clusters) optimization. 
Any partition with classes tagged during static analysis as non offloadable is discarded. 
In Section~\ref{sec:decisions} we discuss how these pre-calculated partition sets are picked for offloading at runtime.



\rone{When it comes to partitioning apps, it only states that it avoids offloading functionality that should execute on the device. It does not describe what the system actually decides to partition. Automated partitioning is one of the contributions of the paper and motivation suggests this can be done in adaptive fashion considering the app and the network. The paper would greatly improve if it contained an adaptive algorithm that intelligently partitions apps. One key challenge in partitioning is that often objects have shared data structure and complex dependencies. For example, parameter settings, environment variables, references are hard to track. So it is non-trivial to capture the whole execution context (but doable).}

\subsection{Offloading Functionality}
\label{sec:offloading}

Once INFv fetches the partition metadata, it needs to provide offloading capabilities and a decision model to ensure that it executes faster and consumes less energy.  




The INFv end device system (Fig.\ref{fig:mobile}) is composed by one or more monitors and one standalone process that provides a network stub and a resource monitoring (RM) system.
Each app process transparently loads and executes an INFv monitor on start.
The App Manager intercepts the app entry points declared in the manifest and binds to the per-device Network Stub. 
The Partition Manager (PM) loads the partitions and instructs the Hook Manager to hook their entry and exit points, enabling the interception of their respective members (methods \& constructors).
Once a target invocation is intercepted, the PM retrieves the current environment-dependent metrics from the RM and decides whether to offload it or not.
If so, a message is created and sent by the AppConnector to the Network Stub service (via IPC) that transparently interacts with the closer network node to execute it.
Network nodes abstract the network topology by providing a message queue (MQ) between the stub and the execution backend.
Within the MQ abstraction, routing is done using the user, device, app and version IDs, along with the fully qualified member name and its arguments.
The Thread Manager keeps a per-thread queue of the offloaded requests and suspends the threads until the invocation result is received or there is an intermediate invocation of a member in the same thread. 

The INFv backend subsystem runs the same monitor functionality over a light-weight app process.
It loads only the required functionality and listens for incoming requests (Request Manager) from the mobile device.
On request it creates class instances and executes their respective methods while managing their state.
Although mobile apps are expected to be short-lived (imposed by screen off events and CPU sleep mode), state has to be kept while there are references to it.
In INFv, remote class instances are replaced locally with ``light-weight'' instances of the same class (bypassing their constructors~\cite{objenesis}), which, on interception, work as proxy objects.
These objects are tracked using a custom weak identity hash map that allows the detection of de-referenced or garbage collected objects, which results in a state invalidation message being sent through the INFv network stub.
App crashes or force quit also trigger state invalidation.
If the crash is INFv's specific, as apps are isolated, the instrumentation can be disabled for the specific app.
Finally, when there are no requests or remaining references the VM can be paused after a certain period. 
More advanced distributed garbage collection mechanisms can be implemented that address some of aforementioned challenges~\cite{abdullahi_garbage_1998}.


While INFv's offloading model resembles Java RMI, not only does Android's Java not support RMI by default, but it generally requires a remote interface (e.g., extending \textit{java.rmi.Remote}).
Refactoring classes is possible at runtime using bytecode manipulation but is either expensive if done at launch time or significantly increases the app size by up to 40\%~\cite{zhang_refactoring_2012} when stored.


\subsubsection{Runtime Decisions}
\label{sec:decisions}

\rone{Also there is a risk that the app might behave incorrectly. How are these handled in INFv? The paper can describe these challenges to highlight its contribution.}

\manote{All INFv related Exceptions extend our own custom Exception. If a non-INFv exception occurs it is passed to the application itself as it should. If the exception is related to INFv (e.g., sudden disconnection), if it occurs on a remote object creation, the object is created locally and it will work as expected using the local resources. If a method invocation over a remote resource fails, in the case of connectivity the invoking thread is paused until it is re-established and, if over a threshold, the user is notified. Other exceptions can drive the classes to be marked as not-to-offload to avoid crashes in future executions. Additionally we describe in the end of this section how the errors in energy measurements can be overcome.}

\rone{After reading 4.2.1 I didn't get how it actually works and how involved the system is. So I'll list some questions I had. What's the underlying technology? Which method calls need to be intercepted? What happens when the partitioning (what to offload) is decided at run time?}

\manote{We had more details in previous submissions, dunno if we should bring them in.}


At runtime the PM decides which of the (offline) pre-calculated partitions should be offloaded.
This decision is based on the partition's (offline) estimated execution parameters (Section~\ref{sec:profiling}) and the periodic device measurements and state monitored by the RM -- network type, bandwidth and latency.
In previous class offloading research~\cite{zhang_refactoring_2012}, a partition would only be valid for offloading if, for each of its classes, all methods perform faster when offloaded.
Unfortunately from our experience this rarely holds, even for CPU intensive classes.
For example, simple methods such as an overridden \textit{toString()} or \textit{equals()}, will in most situations perform worse than local when offloaded due to the RTT.
In INFv, instead of considering methods individually, for each class we aggregate its method's execution and energy consumption based on their observed method invocation frequency ($f_i$).
Therefore, methods that perform better when offloaded can compensate for the least performing methods \emph{if these do not occur too often}.
A class $c$ is valid if: $\sum_{m=1}^{M} f_{m}*t_{m\_local} > \sum_{m=1}^{M} f_{m}*(RTT + (i_{m}+o_{m})/r+t_{m\_offload})$, where $M$ represents the number of class methods and $f_{m}$ the method's invocation count normalized by the total number of invocations observed for all the class methods. 
The size of the method's input and output parameters are represented respectively as $i_{m}$ and $o_{m}$. 
These are divided by the transmission rate ($r$) and, along with the RTT, are only counted for methods interfacing between the local and remote partitions (partition boundary).
Finally the local ($t_{m\_local}$) and remote ($t_{m\_offload}$) execution times are a function of the CPU frequency difference between the mobile device and the node.

Classes are similarly validated based on the energy consumption of their edges, except that a method's energy consumption is only considered for edges that bridge partitions.
For these methods (technically the edges between such classes), only the state transmission costs and the normalized invocation frequency are used when calculating the class's energy consumption.

INFv validates the pre-calculated partitions by increasing the number of partitions ($N$) until it finds a valid partition.
In the worst case scenario $N$ is equal to the number of classes and valid classes are offloaded individually\footnote{In practice, a threshold is selected via modularity optimization~\cite{blondel2008fast} to avoid high link density classes to be executed separately.}. 



Finally, the existing Android diagnosis resources \linebreak(\texttt{dumpsys}) are used to report potential anomalies (e.g., high energy consumption) and processed (offline) by a negative feedback loop to reduce the deviations between estimated and observed values.
An advantage of INFv over app repackaging systems is that it is able to dynamically load new partitions and metadata to handle such cases.

\subsection{Network Subsystem}
\label{sec:network}



The network subsystem abstracts the communication between mobile devices and the executing backend.
It can be deployed on ISP's RANs where the latency to the User Equipment (UE) is minimal (15-45ms~\cite{laner_comparison_2012}).
Offloading requests share a common end-point IP, which can be intercepted directly at the base station~\cite{mec,lipa}, where INFv terminates the traffic and performs further routing.
INFv uses a pub-sub MQ system (as proposed by MEC) to store the processed requests while the routing decisions take place and to perform in-network communication.

Because nodes cache functions, there will be multiple copies in the network. 
The first job of the network subsystem is to route user requests to the closest copy.
Functionality execution consumes both CPU and memory as well as other resources (e.g., bandwidth).
INFv focuses on the first two since they are usually the most dominant resources. 
The second job of the network subsystem is to balance the load of executing functions. 
The goal of load balancing is achieved by strategically dropping or forwarding computation tasks to other nodes to avoid being overloaded. 
However, instead of distributing load uniformly over all available nodes, a service is better executed as close to a client as possible to minimize latency.

\rone{My general feeling is that load balancing may not be the first order concern for mobile code offloading. We have many examples of working systems that are being load balanced. Most of the work in this area assume a small cloud (e.g., single rack) in the access network. It would be good to justify the problem a bit more. The approach also has drawbacks---passing application data between docker instances (applications) incurs latency overhead.  - What is the impact of dropping a offload request?
}
\manote{Package drops will increase the method invocation \"latency\" in the sense that there will be retrials.}

Centralized coordination is not ideal for practical deployment due to three reasons: 1)~A central solver needs global knowledge of a network and maintaining such knowledge up-to-date is costly, 2)~the optimal strategy needs to be calculated periodically given the dynamic nature of network and traffic, and 3)~there is a single point failure. 
Therefore, we study and implement two basic heuristic strategies -- passive \& proactive in INFv.
Both strategies have rather straightforward implementations and try to minimize latency. For the proactive one, we apply a simple $M/M/1$-Processor Sharing (i.e., $M/M/1-PS$) queuing model to estimate the future queue length. The workload on each node can be further estimated based on the predicated queue length and periodically measured CPU and memory consumption of each function. Next we sketch the core idea behind each heuristic; please refer to~\cite{wang_c3po:_2016} for further algorithmic details.

\rthree{The truly novel part should be the one related to operator deployment - however there you design a bunch of algorithms which you don't implement in practice, and you only simulate in completely unrealistic networks (e.g. grid). This leaves many questions open, like: how do I discover my operator's offloading points? What happens when I am roaming? How do the different offloading VMs change stats (required by your protocol)? etc.}

\manote{Is there any study on the mobility of users that shows that users are less likely to use the phone while moving? I mean if you put the phone in your pocket the app is gonna be stopped anyways, you move and when you reconnect all your state will be recreated. Apps are short-lived. Nonetheless, we can probably defend state migration as app state is quite small, we would just need to replicate requests to both machines until we shut the furthest away down.} \liang{yep, you can add a couple of sentences in this section to make the motivation a bit stronger, and discuss a bit about the state migration cost.}


\textbf{Passive Control}: Nodes execute as many requests as possible before being overloaded.
If the node is overloaded, the requests are passed to the next hop along the path to a server, or dropped if the current node is already the last hop in the ISP network. \liang{Or sending it to the server, which is probably more expensive?}

\textbf{Proactive Control}: Nodes execute requests conservatively to avoid being overloaded. 
To do so, a node estimates the request arrival rate to further estimate the potential consumption. 
If the estimate shows that the node may be overloaded, it executes some and forwards the rest to the next hop neighbor with the lightest load. 
NB: This strategy requires exchanging state information within a node's one-hop neighborhood.

\section{Architecture Evaluation}
\label{sec:eval}

\begin{figure} [t!]
\includegraphics[width=8cm]{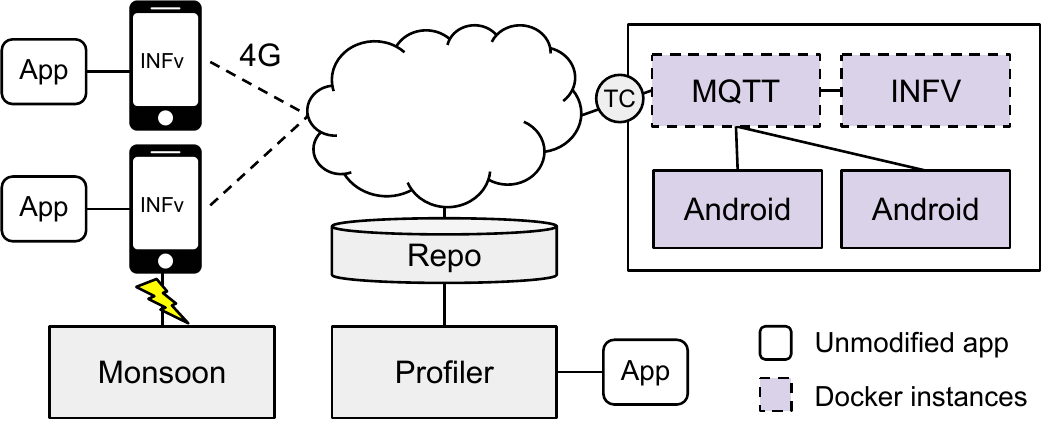}
\caption{Experimental setup.}
\label{fig:last}
\vspace{-5mm}
\end{figure}

We first describe how a typical deployment of INFv then present thorough evaluations demonstrating that INFv delivers on its promise of energy savings and faster app execution.

\textbf{INFvs use case}: \textbf{1)} The profiler analyzes apps and computes partition sets; \textbf{2)} The user equipment (UE) installs apps and the local INFv installation downloads their partition metadata; \textbf{3)} The UE launches apps, and if the network conditions (bandwidth \& latency) are favorable, the execution is offloaded; \textbf{4)} the INFv network system forwards offloading request to an available backend, and finally, \textbf{5)} the backend executes requests.


\textbf{Our experimental setup} is shown in Fig.~\ref{fig:last}.  
Both INFv's network subsystem and the MQ system are run within docker containers.
The selected MQ protocol was MQTT (optimized for mobile devices) and our messages are encoded with protocol buffers~\cite{protobuf}.
The app profiler utilizes hardware-level virtualization (Android x86) to profile the user apps.
Power measurements are taken with a Monsoon Power Monitor and latency is emulated using \texttt{TC NetEm} (Traffic Control Network Emulation). 
In all experiments, phones are factory reset with Android 4.4, no Google account, and INFv pre-installed.
In Section~\ref{sec:evaluation:facedetect} and~\ref{sec:evaluation:linpack} we used a Galaxy S2 (i9100) and offload to an Intel Q6600 (4GB of RAM and a 100 Mbit fiber connection).
Section~\ref{sec:runtimeeval} uses a more up-to-date setup: a Galaxy S5 and an Intel i7 4790K (16GB RAM, 300 Mbit fiber).
Additionally, while 3/4G setups use real mobile networks, in WiFi the UE and backend share a common WiFi access point.


\textbf{The offloaded apps} were \emph{Linpack}~\cite{lp}, \emph{FaceDetect}, \emph{QuickEditor}~\cite{quickeditor}, and \emph{QuickPhoto}~\cite{quickphoto}.
\emph{Linpack} is computationally intensive and \emph{FaceDetect} has high state transmission costs and have been widely used to benchmark MCO performance~\cite{kosta_thinkair:_2012,zhang_refactoring_2012,Shi:2014:CCO:2632951.2632958}.
\emph{QuickEditor} and \emph{QuickPhoto} both use Google Drive, and although not computationally intensive, they exemplify 1)~how INFv can target common functionality across apps and 2)~how INFv can provide functionality otherwise absent from a device.
Each app is standalone, i.e., no client/server counterpart, and has no special design decisions or implementation to facilitate code offloading.

\rthree{The evaluation is completely underwhelming: you select two outliar apps that are obvious offload candidates (linpack is ridiculous if you ask me) and only evaluate those in a simple setup; this part however is, more or less, completely covered by previous work and only shows your system works. The second part is run entirely in simulation and I can't really take anything interesting away from it. This two parts feel almost unrelated, and it doesn't help the paper at all.}
\rone{Evaluation is based on a couple of home grown apps. (It does not show the full capability of the system.)}

\manote{Clearly users are not impressed by our two apps. In our previous paper we add plus a text editor and patch including a security app. Perhaps I can look into one of Tarkoma students \footnote{https://github.com/huberflores} which has a couple of apps used in their paper such as NQueens and MiniMax Chess. They kinda cheated as they basically just did two apps for client and server, no real automation or anything. But at least the chess shows the source algorithm. Regarding the second comment, well this is an extremly complex process and there is a reason why all papers work with opensource apps. If we are to go with a real app, i would prefer opensource and we need to know exactly what to offload.}




\subsection{Impact of Code Partitions}
\label{sec:evaluation:linpack}




\begin{figure} [t!]
\includegraphics[width=0.47\textwidth]{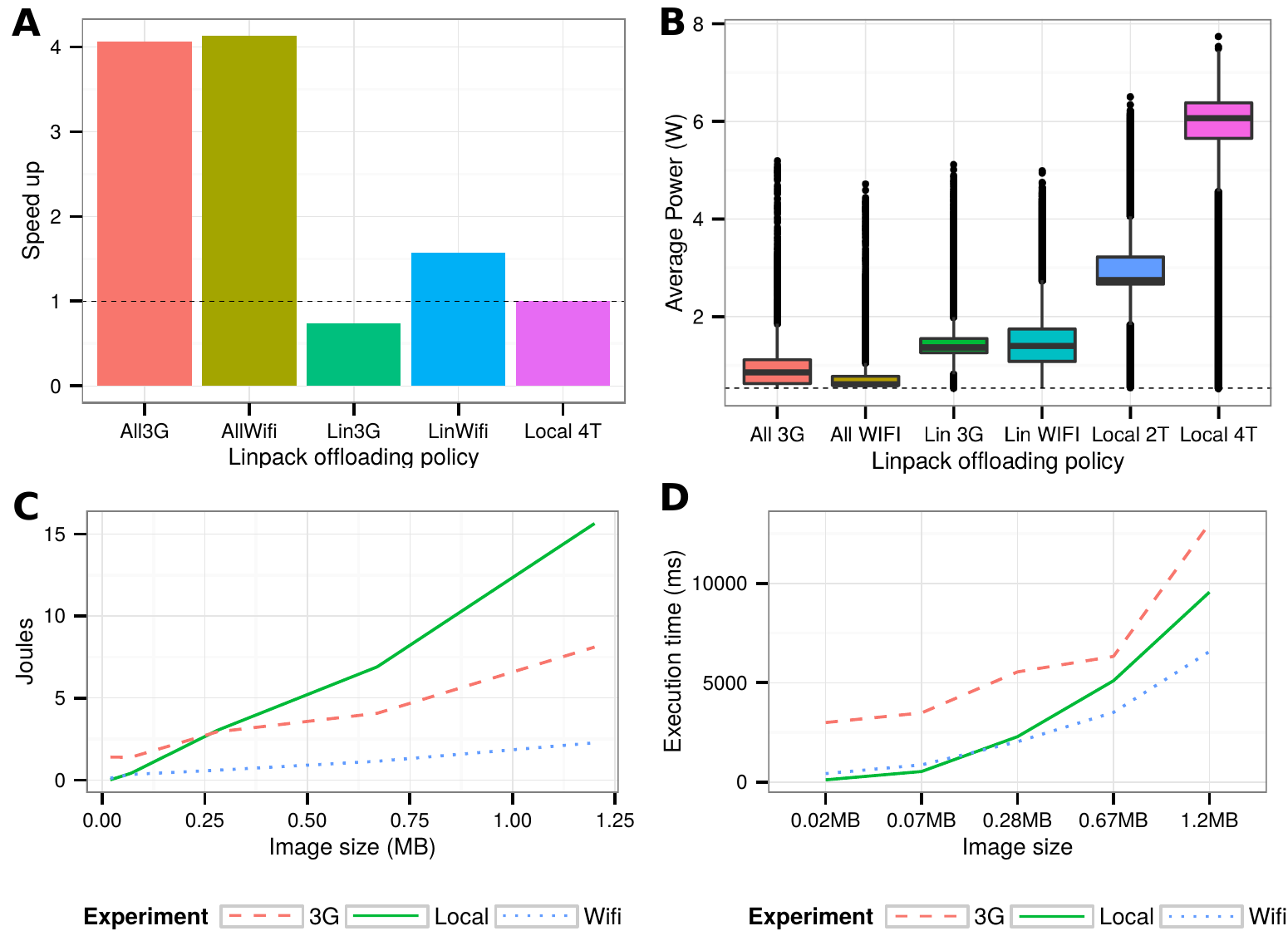}
\caption{Benefits of mobile code offloading for two apps: Linpack and FaceDetect.}
\label{fig:offloadbenefits}
\vspace{-5mm}
\end{figure}

Linpack measures a system's floating point computing power by randomly generating a dense matrix of floating point numbers on $[-1, 1]$ and performing LU decomposition for $M$ cycles (iterations).
We added support for multi-threading, a feature many previous automated offloading architectures do not or only partially support~\cite{cuervo_maui:_2010,chun_clonecloud:_2011,kemp_cuckoo:_2010}.
We experimented using two different offloading partition sets.
For the first (\emph{All}), GN provides two partitions that minimize network communication:
1)~a partition that interacts with the UI, therefore invalid for offloading, and 2)~a valid partition that performs computation, i.e., all computation is performed remotely and there is almost no communication costs as threading occurs on the backend.
The second (\emph{Lin}) represents the worst case scenario by offloading individual classes (i.e. $N$ partitions where $N$ is the number of classes in the app).
For \emph{Lin}, only the Linpack class and calculations are offloaded, and thus, threading, as well as pre- and post-cycle processing, occurs on the client with a higher communication penalty.
For this experiment (and the next) we perform unconditional offloading (i.e. no runtime decisions) to depict the partition trade-offs.

Fig.~\ref{fig:offloadbenefits}A plots the speed up for the Linpack benchmark running 4 threads, showing that INFv provides the expected computational performance benefits of code offloading.
For the \emph{All} partition, INFv achieves a speed up over 4.0x on both WiFi and 3G since there is almost no communication.
When the more restrictive partitioning (\emph{Lin}) is used, the WiFi experiment achieves a 1.57x speed up.
However, the 3G experiment shows \emph{reduced} performance (0.73 speed up), due to the 3G latency and the high frequency of communication (up to 40 messages, 10 per thread, for each cycle).


Next, Fig.~\ref{fig:offloadbenefits}B plots the distribution of power consumption for both partition sets.
The local execution with 2 and 4 threads had a median power consumption of 3 and 6W, respectively; for offloaded executions it was below 2W.
For the WiFi \emph{All} partition, the quartiles show a tight distribution of power consumption and, since the UE was mostly idle with a mean energy consumption close to the observed Android background activity (dashed line), the overall energy consumption was reduced by up to 4 times.

Offloading without modularity optimizing can result in reduced performance in high latency networks despite the power consumption reduction.
Taken together, Fig.~\ref{fig:offloadbenefits}A and~\ref{fig:offloadbenefits}B, show that INFv's MCO provides real benefits but also demonstrates the trade-offs of different partitionings.


\subsection{Cost of State Synchronization}
\label{sec:evaluation:facedetect}

The FaceDetect (FD) app, which finds the coordinates of faces in images, is useful for measuring the trade-offs between data transfer and computational speed up.
The offloaded partition contains the classes interacting with the face detection APIs and the client device just sends the underlying Android Bitmap object and receives an array of coordinates.
The execution time and power were measured from when the app starts up to when the results are drawn on the screen.
To test the impact of data transfer, we use multiple images from the AT\&T face database~\cite{facedb} ranging in size from 0.02MB to 1.2MB.


Fig.~\ref{fig:offloadbenefits}D plots the execution time as a function of image size and we can see that local execution is faster for images $\leq$ 0.07MB.
This is because detecting faces in small images does not have enough computational cost to outweigh the communication costs of offloading.
For larger images, the WiFi communication costs are compensated by the VM processing speed. 
For example, with the 1.2MB image, using WIFI has an execution speed 1.45x faster, but for a 0.2MB image offloading was 3.86x slower. 
Unfortunately offloading was never justified (in terms of execution time) over 3G due to its high latency.

Fig.~\ref{fig:offloadbenefits}C plots the total Joules FaceDetect consumed, not counting baseline OS consumption.
We note that for small images ($\leq$ 0.25MB), local execution results in less power consumption than 3G, although after this point, offloading over 3G saves energy.
Offloading over WiFi results in lower power consumption than local execution for all but the smallest image in the dataset.
For example, an image with 1.2MB consumes 1.9x less battery when offloaded via 3G connectivity and 6.9x less battery if offloaded via WiFi.
In the case of WiFi, the decreased execution time due to the VM processing power is the significant factor in energy savings.
The 3G energy consumption is higher than WiFi for two reasons: 1)~there is additional radio overhead for 3G and 2)~the total execution time is larger due the higher RTT.

The main take away from these experiments is that, as in the Linpack experiments, INFv's offloading engine provides both computational and energy benefits.
However, if there is substantial interaction between local and offloaded objects that involves passing a lot of data, it can result in a net \emph{loss} of performance.
In Section~\ref{sec:runtimeeval} we show how profiling metadata can be used to prevent such scenarios.

\subsection{Responsiveness to Jitter}
\label{sec:jitter}

\manote{we need here some short sentence to connect it with previous experiments}

\begin{figure} [t!]
\includegraphics[width=8cm]{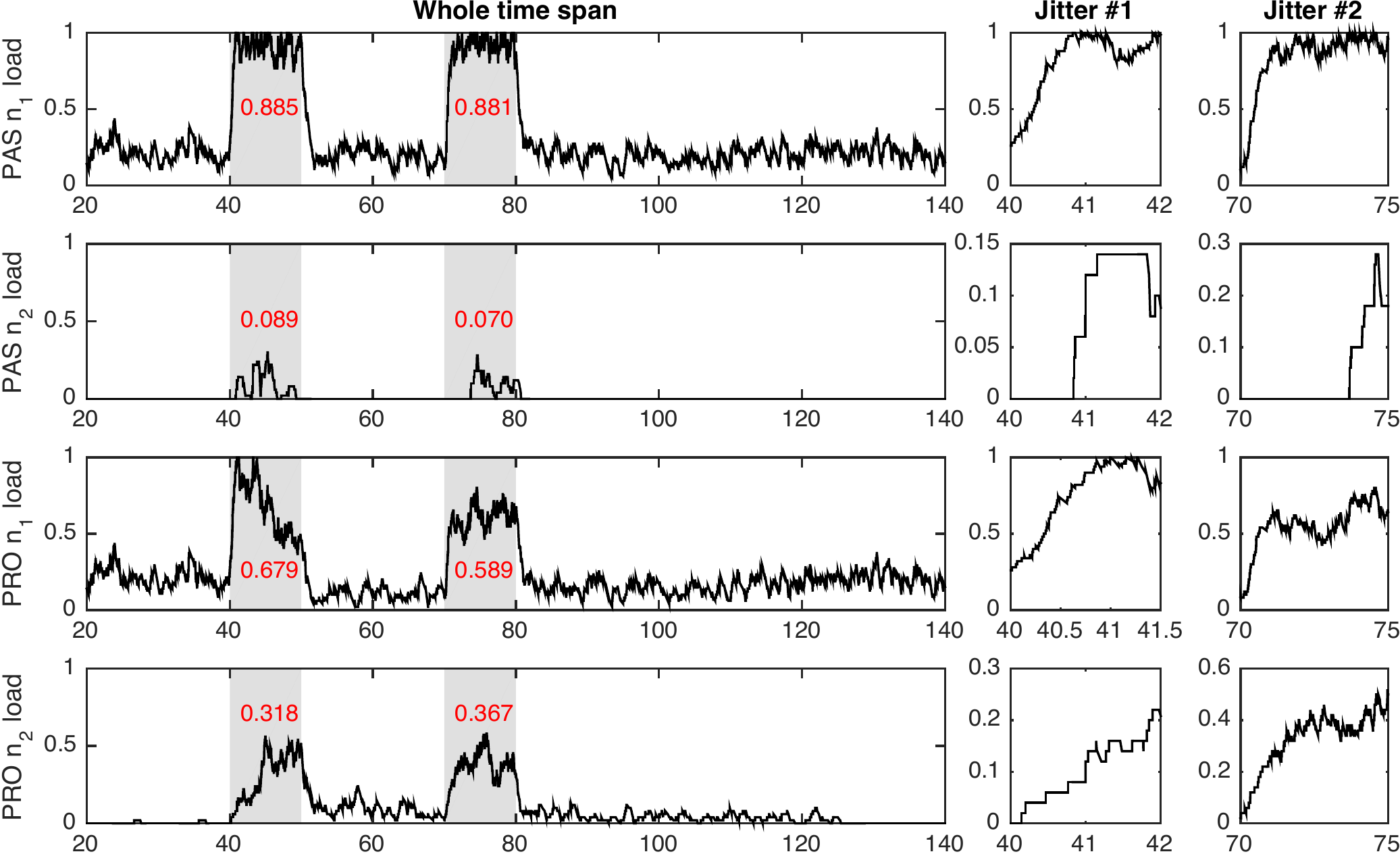}
\caption{Comparison of two control strategies by examining two adjacent routers: client $\rightarrow$ router $n_1$ $\rightarrow$ router $n_2$ $\rightarrow$ server. Two jitter are injected at time 40 ms and 70 ms. $x$-axis is time (ms) and $y$-axis is normalized load. Red numbers represent the average load during a jitter period.}
\label{fig:3}
\vspace{-5mm}
\end{figure}

Next, we study how INFv's control strategies respond to sudden increases in workload (i.e., jitter).
We setup a toy network topology composed of a client, two routers ($n_1$ and $n_2$), and a server (acting as a catch-all for requests not handled by $n_1$ or $n_2$); i.e., client $\rightarrow$ router $n_1$ $\rightarrow$ router $n_2$ $\rightarrow$ server.
We simulate the client's request flow at a stable rate of $\lambda = 1000/s$ but inject two instances of jitter at $6\lambda$ for 10ms at time 40ms ($j_1$) and 70ms ($j_2$).
Fig.~\ref{fig:3} plots the workload over time when the routers use a passive strategy (PAS$n_1$ and PAS$n_2$ in the first two rows) vs. a proactive strategy (PRO$n_1$ and PRO$n_2$ in the second two rows).
The two right most columns zoom in to the period when $j_1$ and $j_2$ have just occurred.

For passive control, PAS$n_1$ takes most of the load (88\%), exhibiting consistent behavior for both $j_1$ and $j_2$.
However, the proactive routers show an interesting variation.
For $j_1$, although PRO$n_1$ successfully offloads 31.8\% of load to PRO$n_2$, it also experiences high load for a period of 2ms (row 3, column 2).
After $j_1$, however, PRO$n_1$ enters a conservative mode.
Thus, when $j_2$ arrives, the load curve on PRO$n_1$ is much flatter with no clear peak appearing at all.
Instead, it proactively offloads more tasks to PRO$n_2$, resulting in PRO$n_2$ absorbing about 36.7\% of the load from $j_2$.
Between 80 and 130ms we see some load still transferred from PRO$n_1$ to PRO$n_2$ because PRO$n_1$ remains in conservative mode.
After 130ms, PRO$n_1$ returns to normal mode and the load on PRO$n_2$ goes to 0.

\begin{figure}[t!]
  \centering
    \includegraphics[width=0.45\textwidth]{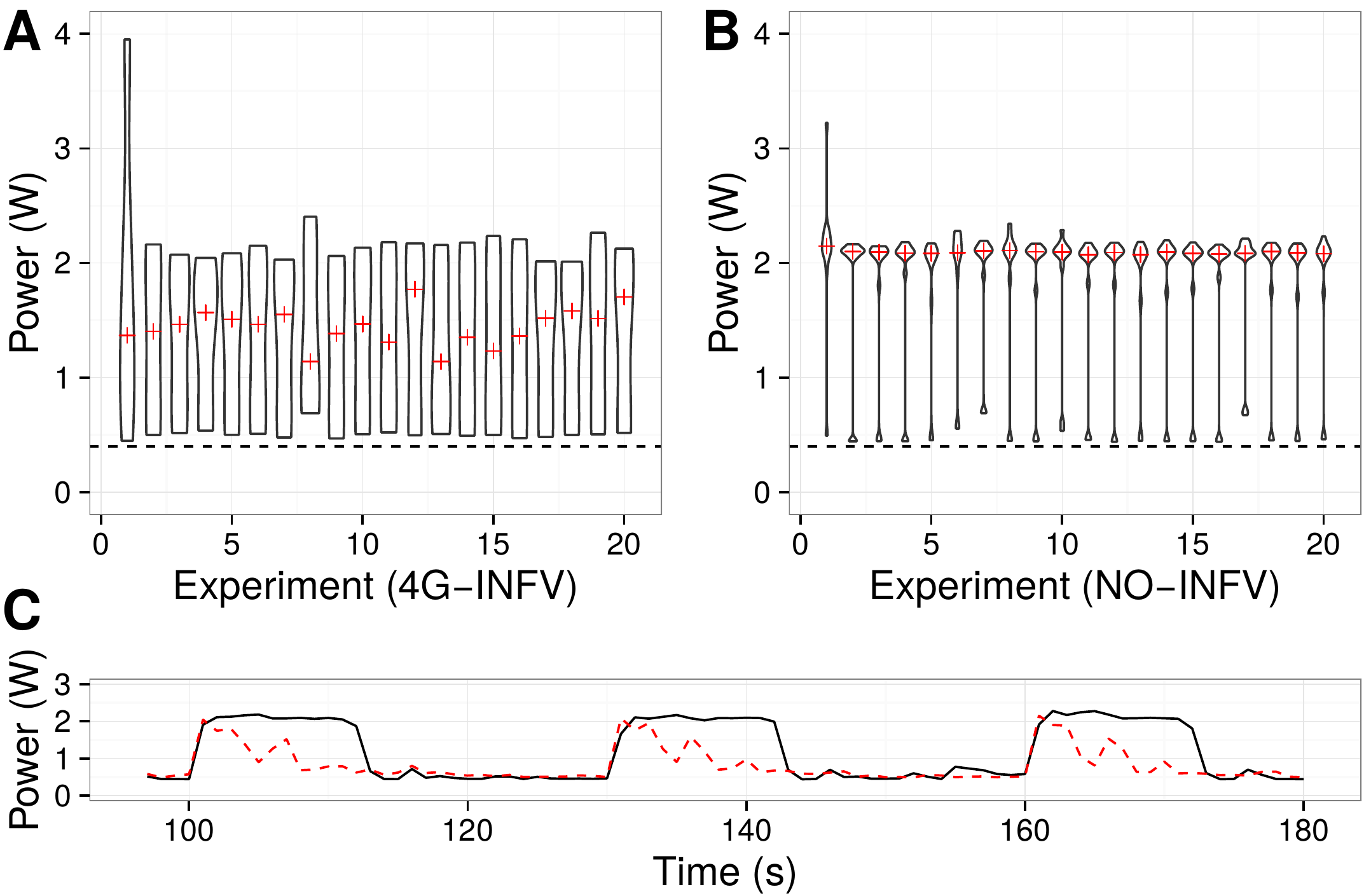}
    \caption{Power consumption distribution (A \& B) for 20 executions over 4G, with and without INFv. Crosses represent the median. In C the dashed line represents INFvs' consumed power versus the local execution (continuous).}
    \label{fig:profile}
    \vspace{-5mm}
\end{figure}

\rone{How accurate is your estimation of energy consumption? It seems particularity difficult given the result of figure 8.A that shows very high variance in the amount of power consumption with INFv.}

By checking the second and third columns, we are able to gain an even better understanding on what actually happens when jitter arrives.
For both $j_1$ and $j_2$, the proactive strategy responds faster; i.e., $n_2$'s load curve rises earlier and faster.
For $j_2$, the proactive strategy responds even faster since PRO$n_2$ is already in conservative mode: PAS$n_2$ only starts taking load at 74 ms, 4 ms later after the $j_2$ arrives at PAS$n_1$.
The major take away here is that INFv is highly responsive to workload jitter due to its network subsystem.



\subsection{Runtime decisions}
\label{sec:runtimeeval}

 \rtwo{It would be good to compare the performance of INFv when offloading is carried out to the cloud rather than ISP. How much latency does it really buy us given the excellent connectivity of large public clouds?}

\begin{figure}[t!]
  \centering
    \includegraphics[width=0.45\textwidth]{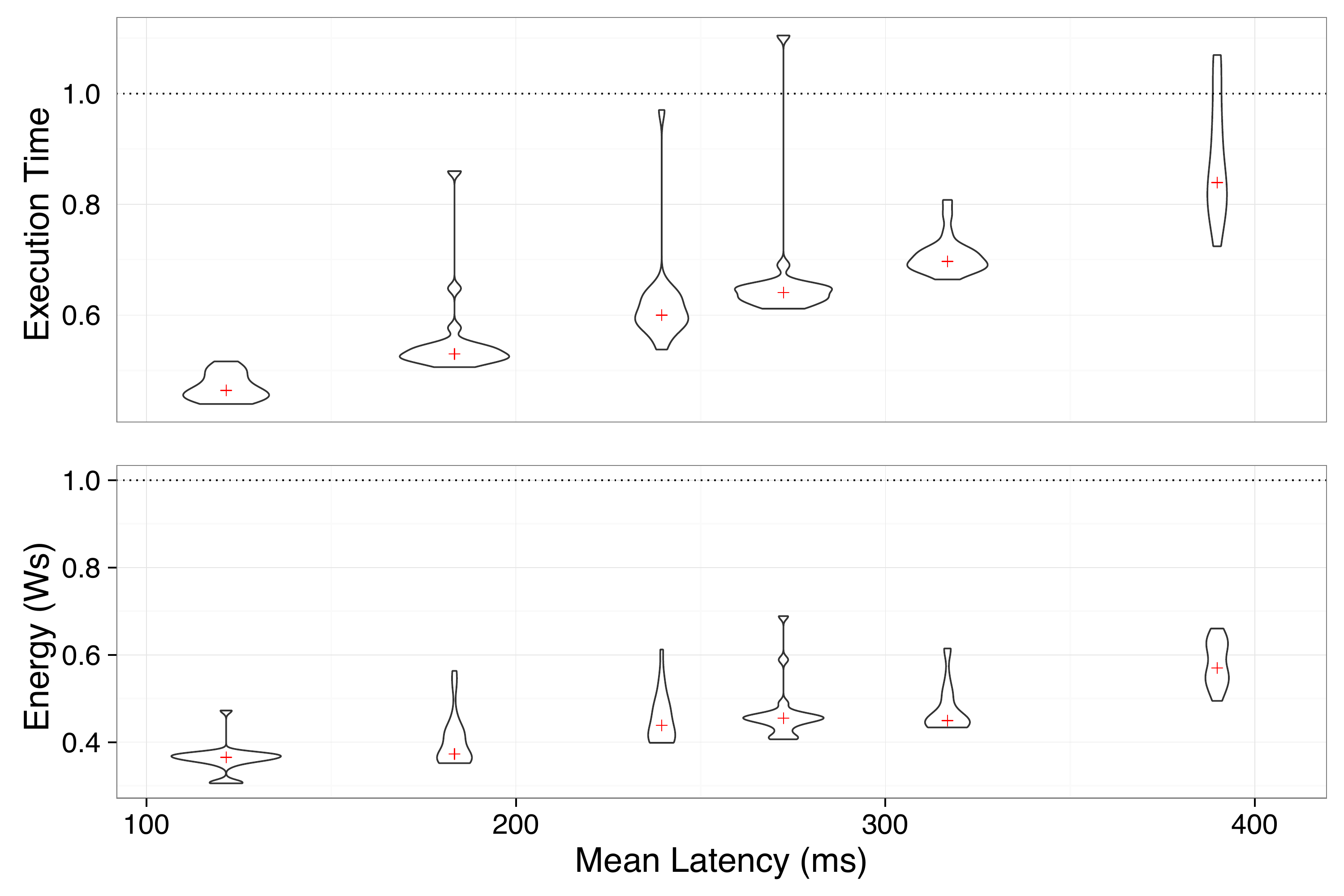}
    \caption{Energy and execution time of FaceDetect execution with INFv enabled. Crosses represent the median. Values are normalized by the mean values of local execution.}
    \label{fig:latency}
    \vspace{-5mm}
\end{figure}




Finally we evaluate how INFv behaves in 4G, with and without additional induced latency, to stress test INFv's runtime decisions.
Fig.~\ref{fig:profile} A) and B) show the observed energy consumption distribution with and without INFv, for 20 experiments using a 1.2MB image~\cite{facesimage}.
In C) we depict the power over time for three of these experiments.
While the execution without INFv (continuous line), consumed over 2 W for most of the experiment time ($\mu \approx 11.57s$), the execution with INFv (dashed line), is mostly comprised of two spikes in energy consumption.
These two spikes represent two distinct phases: 1) image transmission and 2) retrieving and displaying the results; and are dependent on the current connectivity state, e.g., transition to a 4G connected state.
In A) and B) we show the power distribution for the 20 experiments, with and without INFV, respectively.
The power distribution in A) has an higher variance but the majority of the observations are lower than 2W ($\mu \approx 1.35W$ and within a 95\% confidence interval of 0.139W \manote{should we keep the confidence interval or mean is enough?}) and its execution is up to 2,8 times faster ($\mu \approx 2.3$ times) than the execution without INFv, which results in over 66\% energy savings over the 20 executions (idle time excluded).
While UEs are becoming more powerful, so are commodity processors and mobile networks, demonstrated by the execution speed improvements in this experiment compared to the 3G experiments (Section~\ref{sec:evaluation:facedetect}).

Fig.~\ref{fig:latency} shows the impact of latency on execution time and energy consumption of FD over 120 executions.
The observed latency consists of the latency induced on the backend network interface (TC in Fig.~\ref{fig:last}) and the real 4G latency.\footnote{Note that while 300ms might be unusual in 4G, it is quite common in 3G.}
The energy consumption is always lower when offloading, we see a reduction in the savings from close to 70\% less energy (no induced latency) to 40\% due to higher latency.
%
A major takeaway here has to do with in-network vs. cloud deployment: the overall LTE round-trip time (RTT) for offloading to a cloud instance is often over 100ms\footnote{We measured a mean latency of 109.4 and 112.2ms from a mobile device with LTE in Barcelona, Spain to Amazon EC2 regions with the lowest latencies (Frankfurt, \textit{eu-central-1} and Ireland, \textit{eu-west-1}).}; quite high compared to hosting the functionality at the ISP's Radio Access Network (RAN) where devices see only 15-45ms RTT~\cite{laner_comparison_2012}\footnote{we confirmed the lower bound LTE latency values using an USRP transceiver~\cite{B210} and a conservative software-based LTE protocol stack~\cite{openair}.}.
I.e., deploying to the RAN can reduce latency by 58.9\% to 86.7\% (over 90ms difference).
Since our results indicate that a 70ms variation in latency can incur up to 24.5\% and 21.6\% increase in the average execution time and energy consumption, respectively, hosting functionality in-network brings clear benefits.
Further, reducing latency via in-network deployment also increases the set of viable candidate apps for offloading to include those that are particularly latency sensitive (e.g., games).


\begin{figure}
  \centering
    \includegraphics[width=0.35\textwidth]{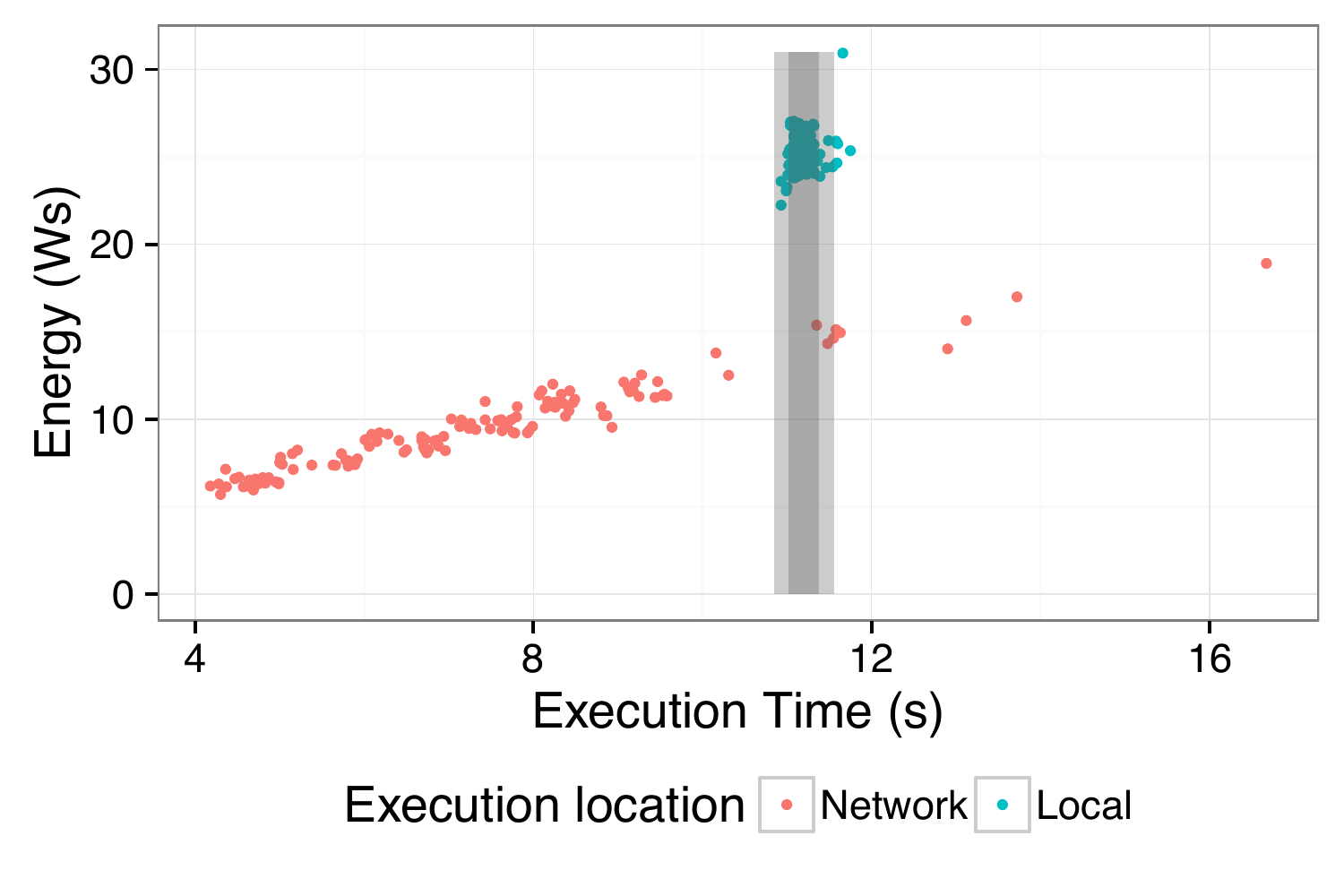}
    \caption{Energy consumption vs. execution time of FaceDetect using INFv under varying latency.}
    \label{fig:lastexp}
    \vspace{-5mm}
\end{figure}

Finally in Fig.~\ref{fig:lastexp} we plot the execution time versus consumed energy for the FD app with the \emph{INFv offloading decision model}.
The RM keeps a rolling window of observed latency (last 3) to the server, which is updated whenever there is no offloading communication (i.e., no threads paused or pending messages) for $>30s$  and the screen is on.
There were 260 experiments over a $>7$ hour period (100ms periods).
We vary the latency (from 0--600ms, with 50ms steps) 30s after each experiment starts. 
INFv decides whether to offload the face detection partition based on the connection properties at runtime (latency and bandwidth vs. transmitted state) and the profiling estimates (i.e., energy and execution time).
Therefore, when executed locally, its behavior should resemble the experiments in Fig.~\ref{fig:profile}A.

Note that the majority (86.7\%) of local execution times fell within one $\sigma$ (the dark gray area in the graph) from those of the experiments in Fig.~\ref{fig:profile}A, and over 96\% of the observed values within two $\sigma$ (light gray).
Moreover, 99\% of local executions' energy consumptions were within one $\sigma$.
\manote{I think this, and the previous observations on the confidence intervals, answer reviewer 1\#12}
There were 4 executions that took \emph{longer} than the local experiments, however, they are an artifact of the periodicity of the  latency measurements: INFv was not able to detect the increase in latency prior to making an offloading decision.
This is important because such cases can occur due to changes in connectivity (e.g., 4G to 3G) or ISP service degradation.
While such impact can be reduced by increasing measurement periodicity, the first is already detected by monitoring changes in the default network interface.

Ultimately, it is clear that, even in the presence of high latency variance, \emph{INFv detected when computation should not be offloaded and energy consumption was greatly reduced for all offloaded experiments}, performing faster than the worst local execution 98.5\% of the time and faster than all local executions 93.2\% of the time.

\subsection{Offloading Popular Libraries and Apps}
\label{sec:offloadable}

\begin{figure} []
\includegraphics[width=8cm]{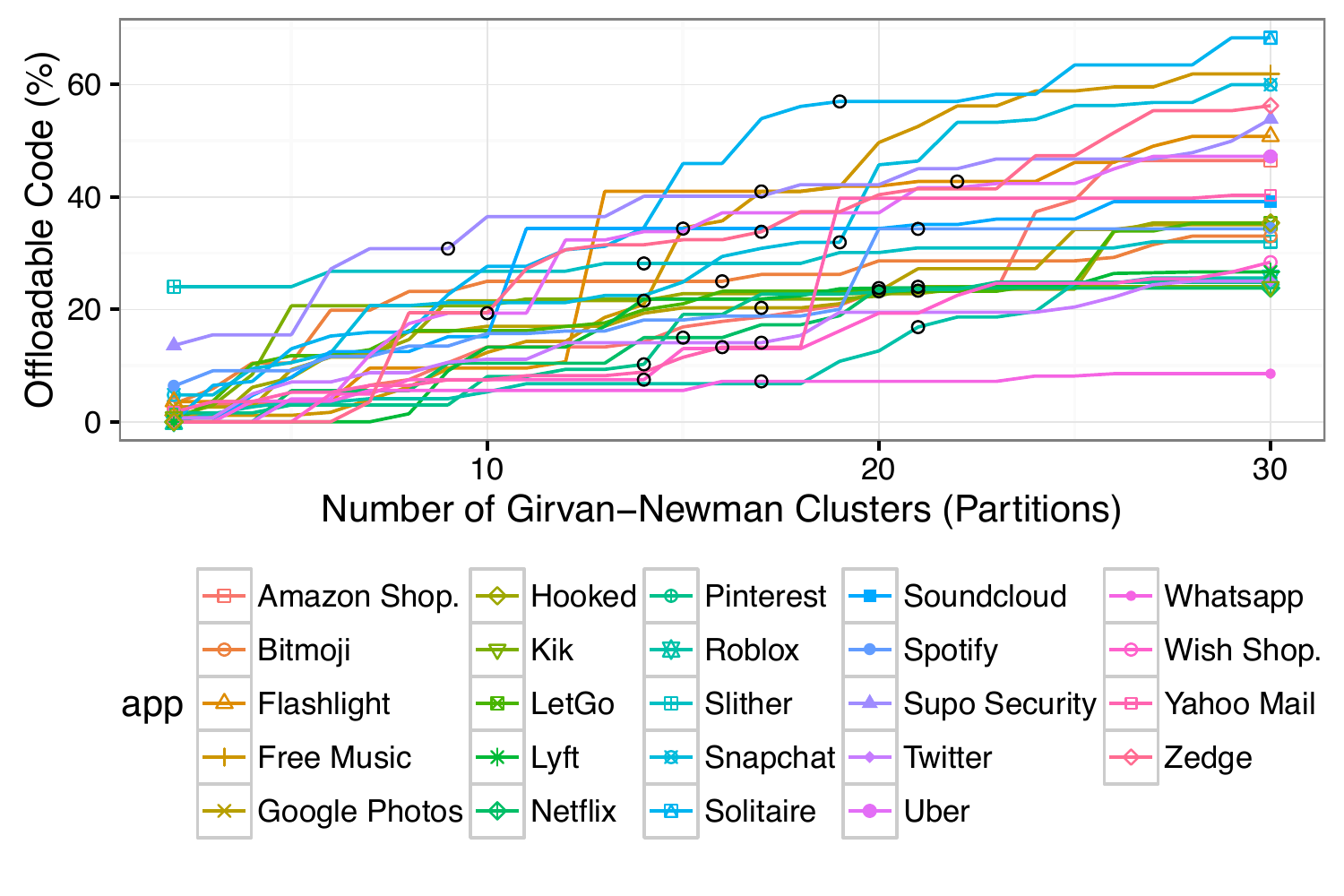}
\caption{Percentage of offloadable code per number of Girvan-Newman partitions. Black circles represent the optimal partition given by the louvain algorithm.}
\label{fig:gn}
\vspace{-5mm}
\end{figure}

To evaluate INFv's support for the most popular libraries and apps, we offloaded an ubiquitous library and performed a partitioning analysis of the top free apps in the GP market.

First, we studied the top 2.5K apps and found that 76.4\% of apps use Google Mobile Services (GMS).
We chose two applications that use a common GMS service -- Google Drive~(GD).
The first, QuickEditor~\cite{quickeditor}, is a text editor that allows users to create, open, and edit text files stored in their GD.
The second, QuickPhoto~\cite{quickphoto}, uses the device's camera to take pictures and upload them to GD.
To support both apps in a device \emph{without GMS installed}, 26 GMS classes were offloaded
, none of them app specific (code available at~\cite{gmsreplace}).
With our in-network solution the number of extra network hops to provide GMS functionality are minimal since GD calls already trigger communication which is forwarded through the RAN.
The network communication overhead is also quite minimal: around 46, 50, and 10B, respectively, to create a class, invoke a method, and receive a response.

Second, to address the concern of how many apps are actually offloadable we inspected the top 24 apps regarding INFv's partitioning and validation mechanisms (Section~\ref{sec:partitions}). 
Our first finding was that only 6\% of app classes extend UI or hardware related classes.
While all other classes could potentially be offloaded, we want to minimize the communication between local and offloaded code.
To this end, we used our dynamic analysis platform to execute the apps for 5 minutes each and extract their runtime call graphs to discover valid offloading partitions (i.e., GN communities).
Previous work~\cite{choudhary_automated_2015} has shown this to be a reasonable interval to achieve high coverage.
Classes that communicate often are likely to share a purpose (e.g., handle UI interaction) and so, building communities based on their communication should separate distinct functionality. 
In fact, Figure~\ref{fig:gn} shows how increasing the number of partitions increases the amount of offloadable code due to this separation of purposes.
At 30 GN communities, all but a single app have between 24\%-68\% of their code suitable for offloading (hundreds to thousands of classes).
If we use the Louvain~\cite{blondel2008fast} algorithm to pick the optimal partitioning (based on modularity), we see find that the median number of partitions per app is 17 and that their offloadable code ranges from 7 to 57\% ($\tilde{x} \approx 24\%$) with only two apps below 10\%.
While the benefits of offloading such partitions are dependent on the runtime environment (e.g., state, network connectivity, etc.) these results indicate that our offloading strategy can be applied to more popular and complex apps with huge real-world user bases.

\section{Deployment considerations}
\label{sec:deploy}

\begin{figure*}[!htb] 
\centering
\subfloat[CDF of package (apk), dalvik executable (dex), and disassembled dex (smali) sizes.]{%
\includegraphics[width=0.3\textwidth]{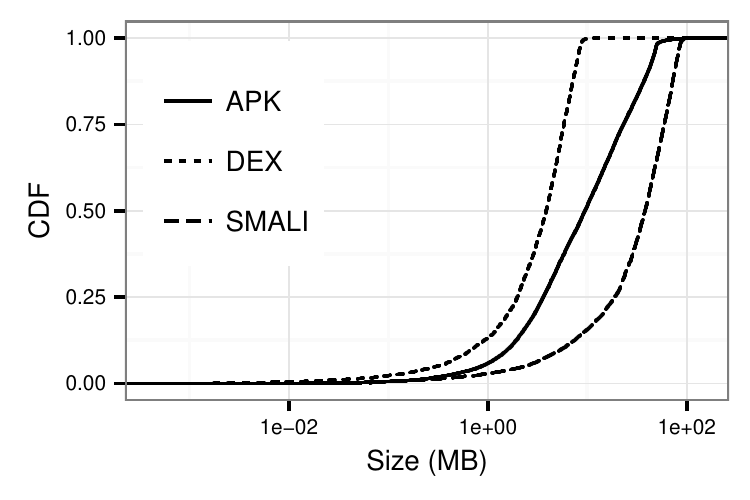}
\label{fig:dex}%
}\hfil
\subfloat[Percentage of unique classes per app as a function of name depth used for comparison.]{%
\includegraphics[width=0.3\textwidth]{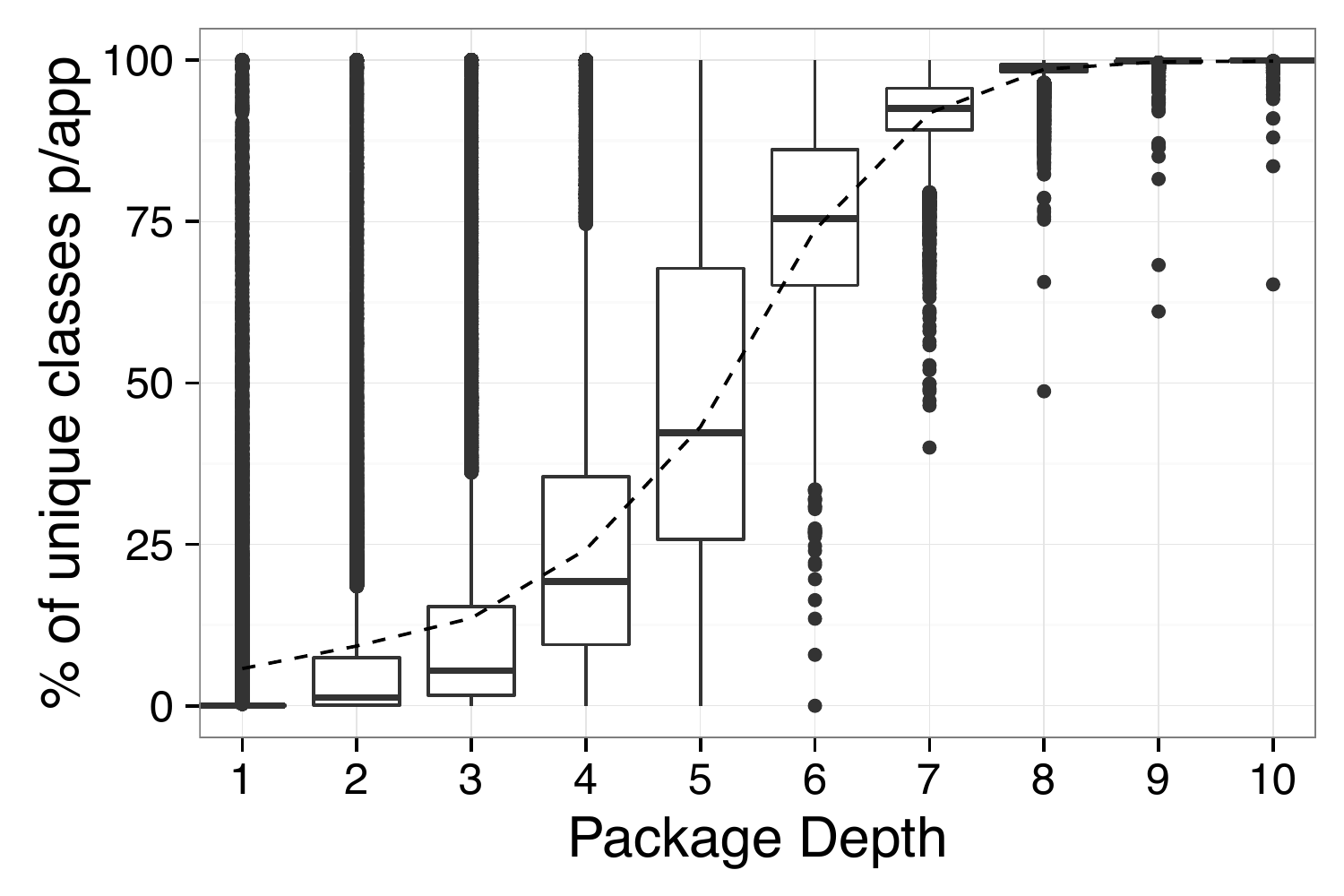}
\label{fig:uniqueclasses}%
}\hfil
\subfloat[Number of classes per app as a function of app popularity.]{%
\includegraphics[width=0.3\textwidth]{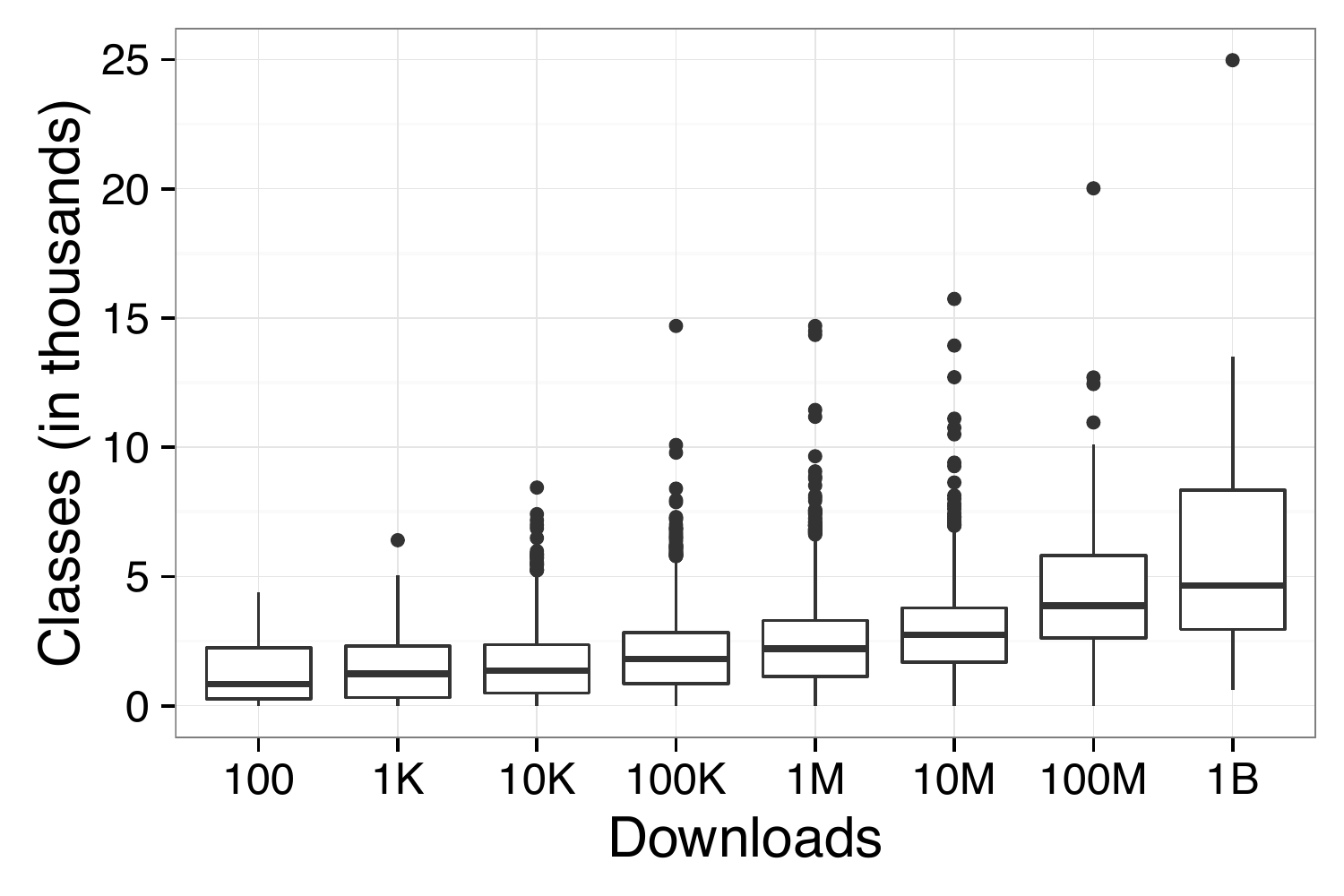}%
\label{fig:popularity}%
}\hfil
\caption{Study of over 20K Google Play apps regarding their size, structure and uniqueness.}
\label{fig:structure}
\vspace{-5mm}
\end{figure*}

\tyson{ is it so important to have heavy overlap of libraries? Presumably, you're going to have classes for each app instance running separately (because each class will likely have different state). Unless you're doing some fancy stateless execution in one JVM, you're gonna have to have multiple processes running - one for each user. This means there is no runtime memory savings from people using the same classes. Obviously, the main benefit is therefore that you you already have the class sitting on the node when a request comes in. Perhaps worth highlighting the point very explicitly.
}

To achieve low latency when serving an offloading request, functionality should optimally be already present in nodes and it should perform well under varying load. This raises two questions: what is the storage cost of hosting the most popular apps?
How well does the network subsystem perform under high-load while reducing the offloading latency?   

\subsection{Storage requirements}
\label{sec:cache}


\rone{Although, the technique in identifying functionality overlap is rough and not very exact, it provides a ball park number. However, I had hard time linking the result/motivation to the rest of the paper. The evaluation does not show that popular classes can be offloaded. Instead, it uses two non-representative workloads. I wasn't sure how the benefit extends to many apps, as implied by the motivation in Section 3.}

\manote{We discussed doing a new plot with X-Axis number of popular apps and Y-axis cache size.}

\manote{Well there should be apps using the face detect apis, so it kind of does offload a common functionality. Perhaps we dont explicitly say its the Android official Face Detect API.}

\manote{Additionally we can try to add a new opensource app.}

\manote{Ack different versions of the libraries can affect caching.}

In this section we give some intuition behind the idea of network functionality caching and empirically show its feasibility via a study of over 20K of the most popular apps on the Google Play Store in February 2016.
As Fig.~\ref{fig:dex} shows, the market app packages (\textit{apk}) are quite small ($\mu \approx 15.3MB$) even considering the actual install size ($\mu \approx 23.9MB$).
Only a fraction of the apk is actually app functionality -- dalvik executable (\textit{dex}) --, which contains the app classes.
For medium/big sized apps (78\% of the apps), when extracted from the \textit{apk}, the \textit{dex} size is on average 34.1\% of the \textit{apk} size ($\tilde{x} \approx 25.3$\%).
Our dataset accounts for over 81\% ($\approx 15,000$ apps \cite{Viennot:2014}) of all Google Play downloads and in total these apps require an aggregated storage of 307~GB (\textit{apk} size), which is a manageable size. 
The large overlap in app functionality can be exploited to intelligently cache functionality in the network.
To understand why and how this is possible, a brief overview of Android app organization and packaging is necessary.
Android apps organize functions into packages which are further identified with a hierarchical naming scheme similar to domain names.
Intuitively, the hierarchical naming could facilitate us in identifying shared functionality across apps.
Unfortunately, naming in Android can be affected by obfuscation, a security mechanism that remaps functionality and package names.
It makes it hard, if not impossible, to detect similarities based on simple name comparisons.
E.g., a package ``\texttt{com.apple}'' from app A and a package ``\texttt{com.google}'' from app B can both be renamed to ``\texttt{a.a}''.
But, since in Android obfuscation (i.e., Proguard) names are attributed alphabetically, we found that it is possible to detect if a package name is obfuscated or not based simply on name length.
We found that 37.5\% of apps have potentially obfuscated packages with name ``a'' (at any given depth), while 82\% of the apps have at least one class named ``a''.
By studying the name distribution with a single character at any given depth, we have found that the majority of obfuscated package names have names between ``\texttt{a}'' and ``\texttt{p}'' (8\% of all packages).
Thus, we \emph{filtered all package names including names with just one character, at any given level.}
Unfortunately obfuscated class names do not follow a similar distribution and such filter would exclude an high number of non-obfuscated classes.
The remaining package names were used to estimate functionality similarity. 
If the same package name exists in two different apps we consider the functionality within this packages to be similar.
Obviously, looking at low depth names ($N < 3$), such as the first depth packages (e.g., ``\texttt{a}'' in ``\texttt{a.b.c}'') which contain all other packages and functionality, many apps are likely to share the same name and therefore, most of the functionality will be considered similar (false positives).
Looking deeper into the package hierarchy, however, can greatly reduce the rate of false positives.
Fig.~\ref{fig:uniqueclasses} shows the percentage of unique classes per app based on a comparison on their first N package names. 
The number of unique classes are calculated as the \emph{total number of classes} in the app minus the number of classes belonging to \emph{non-obfuscated} package names that also exist in at least one other app.
Considering that the package name distribution has $\mu \approx 4.7$, $\tilde{x} \approx 5.0$ and $\sigma \approx 1.6$, and that most package names have a depth between 4 (Q1) and 6 (Q3), even if we do a conservative comparison of packages based on their first 5 name depths (50th percentile), we can see that \emph{only 47\% (mean) of apps' classes are unique}.
Note that while higher values ($N \geq 8$) reduce false positives, they also increase the number of false negatives as classes within packages with smaller depths are considered as unique. 
For a depth of 4, which is likely to include the name of the app and respective developer (e.g., ``\texttt{com.facebook.katana.app}''), 75\% (median) of apps' functionality is common with at least one other app.
The analysis thus indicates that there is a substantial app's functionality overlap in the Android ecosystem.

Extracting the app classes (\textit{classes.dex}), the storage requirements to host over 81\% of the most downloaded apps, are already reduced by 74\% ($\approx 80$GB).
If common functionality is co-located, the total reduction can be up to 93.5\% ($<20$GB, based on the 4th depth median overlap). 
While our app analysis platform requires at least the full \textit{apk} files for analysis, the class overlap provides a unique opportunity for deployment in a modern ISP network, allowing INFv to exploit the network topology and ensure that functionality is available as close to users as possible.


\subsection{Scalability to Workload}
\label{sec:workload}

In this section we perform a large network simulation using a realistic ISP topology (Exodus~\cite{spring_measuring_2002}) with Icarus~\cite{icarus-simutools14}. 
We use a Poisson request stream with $\lambda = 1000/s$ for the arrival rate; increasing the request rate introduces more load to the network.
To simplify the presentation, we assume CPU is the first bottleneck in the system for computationally intensive apps and all experiments are performed $>$50 times to ensure the reported results are representative.

\begin{figure} [t!]
\includegraphics[width=8cm]{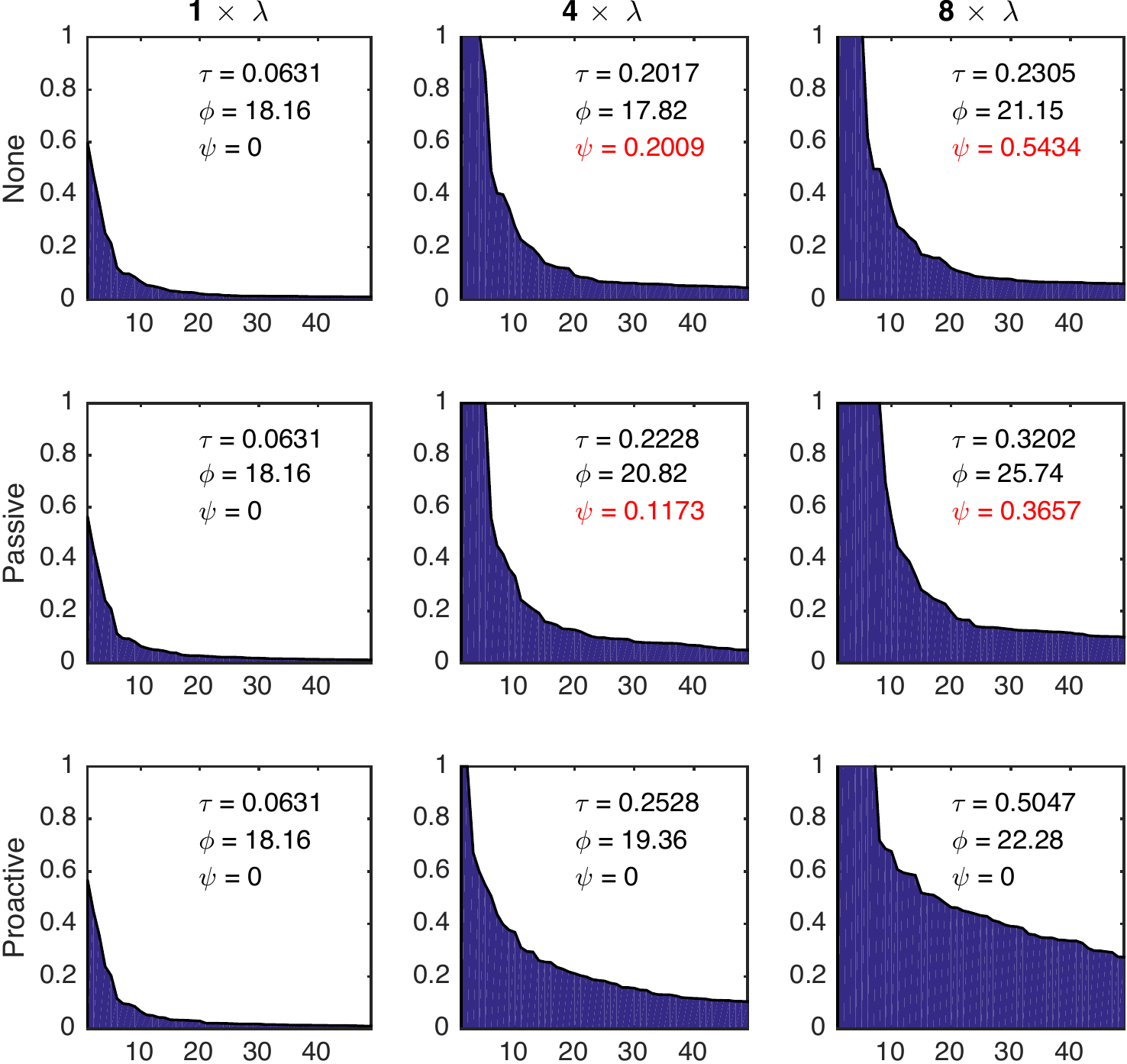}
\caption{Comparison of three strategies (rows) with increasing load (columns). $x$-axis is node index and $y$-axis is load. Top $50$ nodes of the heaviest load are sorted in decreasing order and presented. Notations: $\tau$: average load; $\phi$: average latency (in $ms$); $\psi$: ratio of dropped requests.}
\label{fig:2}
\vspace{-5mm}
\end{figure}

Fig.~\ref{fig:2} shows the results of using three strategies (one for each row) with three workloads (one for each column). There are 375 nodes in the network and we randomly select 100 nodes of degree one as access points to receive user requests. The average load of each node is normalized by its CPU capacity and only the top 50 heaviest loads are presented in a decreasing order. 
By examining the first column, we can see all three strategies have identical behaviors when the network is underutilized with a workload of $\lambda$. The heaviest loaded node uses only about 60\% of its total capacity. However, as we increase the load to $4 \lambda$ and $8 \lambda$, the three strategies exhibit quite different behavior.
The experiment without a control strategy (``none'') at the first row, the figures remain the similar shape. Since no load is distributed and a node simply drops all requests when being overloaded, it leads to over 54\% drop rate with load of $8 \lambda$. 

For passive control (second row), we can see both the heads and tails are fatter than ``none'' control, indicating that more load is absorbed by the network and distributed on different routers. This can also be verified by checking the average load in the figure: given a load of $8 \lambda$, passive control increases the average load of the network from $0.2305$ to $0.3202$ compared to using ``none'' control. However, there is still over $36\%$ requests dropped at the last hop router. This can be explained by the well-known small-world effect which makes the network diameter short, so there are only limited resources along a random path.

Among all the experiments, a network with proactive control always absorbs all the load, leading to the highest average load in the network which further indicates the highest utilization rate. As the workload increases from $\lambda$ to $8 \lambda$, average load also increases accordingly with the same factor. One very distinct characteristic that can be easily noticed in the last two figures on the third row is that the load distribution has a very heavy tail. This is attributed to the proactive strategy's capability of offloading  to its neighbors. It is also worth pointing out that we only measured the latency of those successfully executed functions, which further explains why ``none'' control has the smallest latency, since offloaded functionality gets executed immediately at an edge router connected to a client, but \emph{more than half the requests are simply dropped and not counted at all}. Comparing to the passive strategy, the proactive strategy achieves lower latency. Further investigation on other ISP topologies shows that latency reduction improves with larger networks.


\section{Conclusion \& Future work}
\label{sec:discussion}

Battery is a huge constraint for mobile devices and the ever growing demands of computation on limited capacity are unlikely to disappear any time soon.
Meanwhile, in-network storage and computation resources are growing.
We proposed INFv to exploit in-network resources for mobile function offloading.
We described its data-driven design and implementation based on a large scale analysis of a real app market. 
Our evaluation demonstrates that INFv's non-intrusive offloading technique can significantly improve mobile device's performance (up to 6.9x energy reduction and 4x faster) and effectively execute functionality in the network while reducing latency.
Our analysis shows the potential for functionality caching and popular app offloading, while also providing interesting insights into Android apps' obfuscation and composition.
INFv is a working system and many of its components are open-sourced~\cite{broker,droidsmali,lp}.

There are some limitations and caveats which deserve further investigation in the future.
First, INFv requires attention to the security and privacy of communication and offloaded functionality.
While it provides isolation and detects the use of critical OS APIs, in the future we will consider techniques for detecting vulnerabilities/malware~\cite{mariconti2016mamadroid, 8514191} and access to privacy sensitive information~\cite{enck_taintdroid:_2014}. 
Although it does not require a custom OS, a one-time root is required, which can be disabled after install.
If deployed by an ISP, it can be pre-installed on devices or installed in stores.
For other scenarios, either root would be required or, an existing vulnerability could be leveraged to install INFv and secure the device~\cite{mulliner_patchdroid:_2013}.


Second, UI automation might not cover all app code and its generated state might not be representative of real user's input.
A significant amount of work exists on improving coverage~\cite{machiry_dynodroid:_2013,mahmood_evodroid:_2014, choudhary_automated_2015, chimp} and we are looking to further improve our dynamic analysis by exploiting crowdsourcing platforms~\cite{crowdflower} to test apps with real users.
Additionally, an iOS implementation should be possible using similar interception mechanisms~\cite{cydia} and static analysis can be accomplished by dumping the decrypted apps from memory~\cite{dumpdecrypted}. We also plan to explore different interception techniques to better support native code~\cite{cydia}.
Finally, we are working to deploy a small-scale real-user test in the coming year to gain valuable feedback to further improve INFv.

\bibliographystyle{abbrvnat}
\bibliography{references}

\end{document}